\documentclass[aps,prb,twocolumn,superscriptaddress]{revtex4-2}

\newcommand{\1}{\text{X}\textsubscript{1}}
\newcommand{\2}{\text{X}\textsubscript{2}}
\newcommand{\3}{\text{X}\textsubscript{3}}
\newcommand{\4}{\text{X}\textsubscript{4}}
\newcommand{\5}{\text{X}\textsubscript{5}}
\newcommand{\6}{\text{X}\textsubscript{6}}

\newcommand{\s}{{s}}
\newcommand{\p}{{p}}

\usepackage{graphicx}
\usepackage{dcolumn}
\usepackage{bm}
\usepackage{amsmath}
\usepackage{amssymb}
\usepackage{xcolor}
\usepackage [english]{babel}
\usepackage [autostyle, english = american]{csquotes}
\MakeOuterQuote{"}
\usepackage{placeins}
\usepackage{soul}
\usepackage{hyperref} 
\hypersetup{colorlinks=true,citecolor=blue, urlcolor=blue, linkcolor = blue}

\begin{document}

\title{\textbf{Polarization-Resolved Core Exciton Dynamics in LiF Using Attosecond Transient Absorption Spectroscopy}
}

\author{Kylie J. Gannan}
 \email[]{kylie\_gannan@berkeley.edu}
\author{Lauren B. Drescher}
 \altaffiliation[Present address: ]{Max-Born-Institut f\"ur nichtlineare Optik und Kurzzeitspektroskopie, Max-Born-Stra{\ss}e 2A, 12489 Berlin, Germany}
\author{Rafael Quintero-Bermudez}
\affiliation{Department of Chemistry, University of California, Berkeley, California 94720, USA}

\author{Navdeep Rana}
\affiliation{Department of Physics and Astronomy, Louisiana State University, Baton Rouge, Louisiana 70803-4001, USA}

\author{Chengye Huang}
\affiliation{Department of Chemistry, University of California, Berkeley, California 94720, USA}

\author{Kenneth Schafer}
\author{Mette B. Gaarde}
\affiliation{Department of Physics and Astronomy, Louisiana State University, Baton Rouge, Louisiana 70803-4001, USA}

\author{Stephen R. Leone}
 \email[]{srl@berkeley.edu}
\affiliation{Department of Chemistry, University of California, Berkeley, California 94720, USA}
\affiliation{Department of Physics, University of California, Berkeley, California 94720, USA}
\affiliation{Chemical Sciences Division, Lawrence Berkeley National Laboratory, Berkeley, California 94720, USA}


\begin{abstract}
The ability to control absorption by modifying the polarization of light presents an exciting opportunity to experimentally determine the orbital alignment of absorption features. Here, attosecond extreme ultraviolet (XUV) transient absorption spectroscopy is used to investigate the polarization dependence of core exciton dynamics in LiF thin films at the Li\textsuperscript{+} K edge. XUV pulses excite electrons from the Li 1\s\ core level into the conduction band, allowing for the formation of a \p-orbital-like core exciton, aligned along the XUV light polarization axis. A sub-5\,fs near-infrared (NIR) probe pulse then arrives at variable time delays, perturbing the XUV-excited states and allowing the coherence decay of the core exciton to be mapped. The coherence lifetimes are found to be $\approx$2.4\,\textpm\,0.4\,fs, which is attributed to a phonon-mediated dephasing mechanism as in previous core exciton studies. The differential absorption features are also shown to be sensitive to the relative polarization of the XUV and NIR fields. The parallel NIR probe induces couplings between the initial XUV-excited \p-like bright exciton and \s-like dark excitons. When crossed pump and probe polarizations are used, the coupling between the bright and dark states is no longer dipole-allowed, and the transient absorption signal associated with the coupling is suppressed by approximately 90\%. This interpretation is supported by simulations of a few-level model system, as well as analysis of the calculated band structure. The results indicate that laser polarization can serve as a powerful experimental tool for exploring the orbital alignment of core excitonic states in solid-state materials. 
\end{abstract}


\maketitle

\section{Introduction}
The absorption of light is one of the most fundamental tools available for investigating the properties of matter. The orbital alignment of excited states is one such property that has been the subject of many investigations \cite{bussert_effect_1987,driessen_alignment_1991,driessen_relative_1991,driessen_n-vector_1992,smith_laser_1992,de_vivieriedle_threevector_1993,smith_initial_1993,spain_orbital_1995,zeidler_controlling_2005,reduzzi_polarization_2015}. These studies involved the preparation of an initially aligned excited population in order to probe alignment-dependent effects of excited states. For example, the alignment of a prepared ensemble has been shown to exert a significant influence over energy transfer rates between excited states upon collision with rare gas partners \cite{bussert_effect_1987,driessen_alignment_1991,driessen_n-vector_1992,driessen_relative_1991,smith_laser_1992,de_vivieriedle_threevector_1993,spain_orbital_1995} and on ionization probabilities \cite{zeidler_controlling_2005}. Recently, attosecond transient absorption measurements probed the alignment of optically forbidden (dark) states in helium \cite{reduzzi_polarization_2015}. Near-infrared (NIR) induced couplings between the initially aligned He \textit{n}p Rydberg states prepared by an extreme ultraviolet (XUV) pulse and dark \textit{n}s Rydberg states that are allowed when the polarizations of the XUV and NIR beams are parallel were eliminated when using crossed polarizations, while couplings between \p- and d-aligned states were preserved. Similar results were obtained in measurements of argon \cite{chew_attosecond_2018}. Considering the complex density of states of many solid-state systems, such sensitivity poses the exciting possibility of an experimental probe to identify the orbital configuration of dark states and of core exciton states. To evaluate the application of this method to the condensed phase, we investigate the polarization dependence of XUV transient absorption of lithium fluoride core excitons.\par
Excitons, charge-neutral quasiparticles formed by bound electron-hole pairs, have been the subject of countless investigations in recent years. Valence excitons, in which the hole lies in the valence band, have been of particular interest for their potential applications in devices, including light emitting diodes \cite{pandey_ultrahigh_2023}, photovoltaics \cite{gregg_excitonic_2003,menke_exciton_2014,classen_role_2020,zhu_exciton_2022}, and quantum computing \cite{ghosh_quantum_2020,harankahage_quantum_2021}. By comparison, core excitons (electron-core hole pairs) have been studied far less extensively, largely due to the high photon energies required for their formation. These form most prominently in ionic insulators, where the poor dielectric screening of the cationic core hole allows for very strong binding of the electron \cite{rubloff_far-ultraviolet_1972,pantelides_electronic_1975}. Early extreme ultraviolet studies noted that core excitons generally produce distinctive, strong absorption peaks \cite{rubloff_far-ultraviolet_1972,pantelides_electronic_1975,haensel_measurement_1968,gudat_core_1974,pantelides_new_1974,kowalczyk_x-ray_1974}. Despite being strongly bound by up to a few eV \cite{stott_core_1984}, the bandwidths of these peaks suggest core exciton coherence lifetimes of only a few femtoseconds. The source of this rapid decay has been the subject of debate for many years, with Auger-Meitner decay processes \cite{citrin_interatomic_1973,lapeyre_photoemission_1974} and phonon-mediated dephasing mechanisms \cite{matthew_breadths_1974,citrin_phonon_1974,mahan_emission_1977} being the leading theories. Time-resolved experimental investigations into the mechanisms were not possible until the development of techniques with attosecond temporal resolution.\par 
In recent years, attosecond transient absorption \cite{moulet_soft_2017,chang_ws2,quintero-bermudez_deciphering_2024}, reflectivity \cite{geneaux_mgo, lucchini_unravelling_2021}, and four-wave mixing \cite{gaynor_nacl} spectroscopies have been employed to investigate the mechanisms driving the short core exciton coherence lifetimes. These have also observed that NIR-induced couplings between bright (allowed) and dark (forbidden) core exciton states can occur \cite{moulet_soft_2017,quintero-bermudez_deciphering_2024,geneaux_mgo,lucchini_unravelling_2021,gaynor_nacl}. This implies that the orbital alignment of the initially excited core-level exciton in the solid can play a role in one-photon absorption to these neighboring states. No previous work, however, has explored the polarization dependence of these couplings as has been done in the preparation and study of aligned orbital states for many isolated atomic systems. LiF, often considered the prototypical ionic insulator, is an excellent system to begin investigations into polarization effects in core exciton systems, as its orbital composition is simple compared to many other materials. As in He, the Li\textsuperscript{+} cation has the 1\s\textsuperscript{2} ground state configuration. A linearly polarized XUV pulse excites electrons from the 1\s\ orbital into aligned 2\p-like core exciton states. Unlike in He, the atomic Li\textsuperscript{+} 3d orbitals lie approximately 7\,eV above the 2\p\ orbitals \cite{moore_atomic_1949}, an energy gap too high to be readily accessible from the 2\p\ bright state with 800\,nm NIR pulses. This greatly simplifies the number and type of couplings possible from the prepared 1\s2\p\ core exciton states, as the NIR probe pulse allows primarily for couplings to nearby \s-like states. Furthermore, the core exciton peaks in LiF are exceptionally strong \cite{haensel_measurement_1968,gudat_core_1974,stott_core_1984}, suggesting that any changes in absorption are expected to produce large transient signals, as observed in other core exciton systems \cite{quintero-bermudez_deciphering_2024,geneaux_mgo}.\par
In this report, the core exciton dynamics of LiF at the Li\textsuperscript{+} K edge are investigated through attosecond XUV transient absorption spectroscopy. Several distinct features are first identified in the linear XUV absorption profile, the nature of which are explored through the experimental transient absorption results and theoretical calculations. The transient absorption spectrum shows several time-dependent features across the 59-72\,eV range, all of which are found to decohere within approximately 2\,fs following excitation. The rapid decay is attributed to a phonon-mediated dephasing mechanism, as in other core exciton studies \cite{moulet_soft_2017,quintero-bermudez_deciphering_2024,geneaux_mgo}. By modifying the linear polarization of the NIR field with respect to the initial linearly polarized XUV field, the transient absorption features associated with couplings between bright and dark excitonic states are suppressed by up to 90\%, thereby revealing the presence of a neighboring resonant dark state and identifying its primary orbital character as \s-like experimentally. Simulations of a model few-level system provide support for the interpretation that the large differential absorption features observed are indeed the result of strong couplings between resonant neighboring excitonic states of selected orbital character. Density functional theory (DFT) and Bethe-Salpeter equation (BSE) computations are presented to provide greater insight into the orbital composition of the structures forming the linear absorption profile, revealing substantial excitonic contributions across the 60-72\,eV range. 
\section{Methods}
A sub-5\,fs near-infrared (NIR) probe pulse spanning 450-950\,nm with 500\,Hz repetition rate is incident on a 20\,nm thick polycrystalline LiF film at variable time delays relative to a colinear extreme ultraviolet (XUV) pump pulse. The XUV pulses are generated through high harmonic generation (HHG) in an argon gas cell using a sub-5\,fs NIR driving field, producing a 20-72\,eV XUV bandwidth. The XUV pump excites electrons from the Li\textsuperscript{+} 1\s\ core level into the conduction band, forming core excitons. The NIR probe then perturbs the XUV-induced polarization of the medium, resulting in a delay-dependent modulation of the XUV absorption profile. Positive time delays are chosen to indicate that the XUV pump arrives at the sample before the NIR probe. The transmitted XUV at various NIR delays is spectrally dispersed by a grating onto an X-ray CCD camera. The large LiF band gap of 13.6\,eV ensures the NIR probe does not excite transitions between the valence and conduction bands of the sample. More details are presented in Appendix \ref{AppA}.
\section{Results}
\FloatBarrier
\subsection{Transient absorption spectroscopy of core excitons in LiF}\label{Experiment}
\begin{figure}
\includegraphics{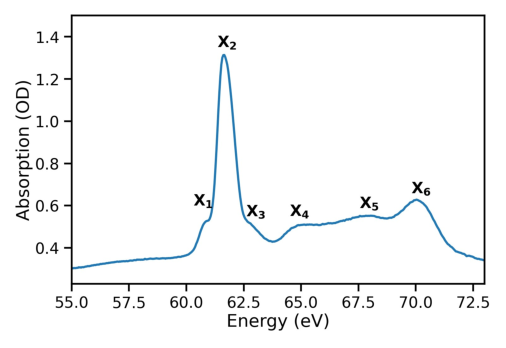}%
\caption{\label{Static} Linear extreme ultraviolet (XUV) absorption spectrum of LiF at the Li K edge (blue curve). Labels \1-\6 highlight transitions from the Li 1\s\ core level. \2 is the 1\s2\p\ bright exciton. \1 is close to the 1\s2\s\ dark state energy. Identification of \1 as the 1\s2\s\ dark exciton has been done previously and is confirmed by calculations here \cite{pantelides_electronic_1975, pantelides_new_1974,kunz_absorption_1973,sonntag_observations_1974,fields_electronic_1977,olovsson_near-edge_2009}. The nature of \3 has not been previously explored in the literature. \4 and \5 lie above the conduction band onset but have been speculated to be excitonic in nature previously \cite{pantelides_electronic_1975}. \6 has been described as an electronic polaron \cite{kunz_role_1972}, although this assignment has been challenged \cite{pantelides_electronic_1975,rottke_probing_2022}.}
\end{figure}

The linear XUV absorption spectrum of a 20 nm thick polycrystalline LiF film at the Li K edge is shown in Fig. \ref{Static}.
The sharp peak at 61.7\,eV (labeled \2) has been attributed to a core exciton formed by excitation of a Li\textsuperscript{+} 1\s\ core electron into Li 2\p-like regions below the conduction band (CB) \cite{haensel_measurement_1968,gudat_core_1974}. Density functional theory (DFT) calculations detailed in Section \ref{Theory} indicate that this peak is composed of three Li 1\s2\p-like excitonic states. In the remainder of this text, Li\textsuperscript{+} and F\textsuperscript{-} are denoted simply as Li and F. A small shoulder on this large peak, observed at 60.8\,eV (\1), has been identified as an optically forbidden (dark) exciton formed between the Li 1\s\ core hole and electrons at the CB minimum, which is of predominantly Li 2\s\ character \cite{pantelides_electronic_1975,pantelides_new_1974,kunz_absorption_1973,sonntag_observations_1974,fields_electronic_1977,olovsson_near-edge_2009}. Observation of this dipole forbidden exciton in the linear absorption spectrum is understood to be the result of a breakdown in symmetry caused by lattice vibrations \cite{olovsson_near-edge_2009}. \6 (70.05\,eV) was originally described as an electronic polaron \cite{kunz_role_1972}, although this description has been debated \cite{pantelides_electronic_1975,rottke_probing_2022}. We offer the alternative explanation that this peak is a 1\s3\p-like excitonic state. We find no discussions of the nature of the \3 peak in the literature, and few descriptions of the \4 and \5 peaks. Pantelides and Brown \cite{pantelides_new_1974} and Pantelides \cite{pantelides_electronic_1975} assert that \4 and \5 may also be excitonic in nature, and that this is possible due to the poorly shielded nature of the Li 1\s\ core hole. We examine the character of \3-\6 more closely using the transient absorption spectra and theoretical calculations presented below.\par
\begin{figure}
\includegraphics{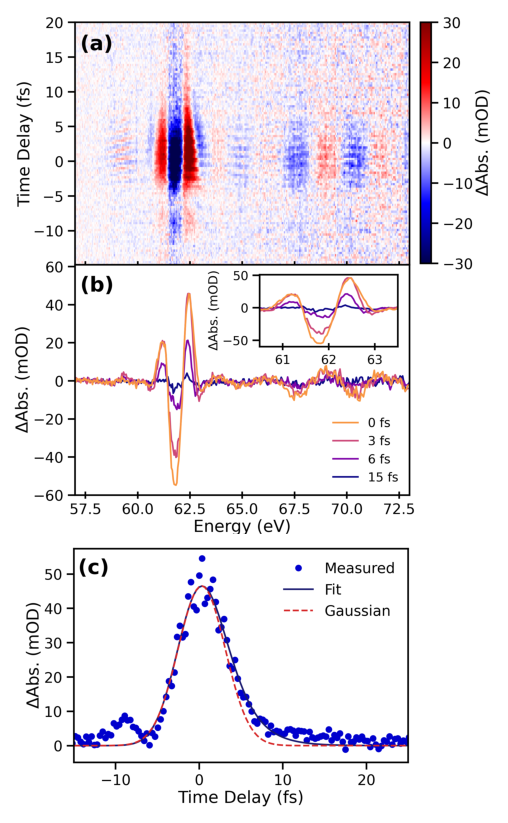}%
\caption{\label{Stabi} Attosecond dynamics of core excitons in LiF observed using XUV transient absorption spectroscopy. (a) Time-delay-dependent transient absorption. (b) Lineouts of delay-dependent transient absorption taken at 0, 3, 6, and 15\,fs. The inset highlights the 60-64\,eV window, showing the subtle delay-dependent shift of the 62.5\,eV peak. (c) Temporal dynamics (blue dots, inverted, at 61.7\,eV) fitted with an exponentially modified Gaussian function (solid dark blue curve). An example Gaussian function (red dashed curve) is plotted to make the exponential tail of the fit clearer. A NIR prepulse evident in the transient absorption spectrum is observed at -9\,fs.}
\end{figure}

The LiF XUV transient absorption spectrum collected with a 6.8\,x\,10\textsuperscript{12}\,W/cm\textsuperscript{2} NIR intensity is shown in Fig. \ref{Stabi}(a).
In this measurement, the temporal delay between the XUV pump and NIR probe with parallel polarizations is stabilized using a feedback signal from the spatial interference pattern of a copropagating 473\,nm continuous-wave diode laser \cite{zinchenko_apparatus_2023}, resulting in a 200\,as standard deviation. The delay range of -22 to +30\,fs was scanned using 0.33\,fs time steps. Fig. \ref{Stabi}(a) shows a reduced temporal axis to better view the core exciton dynamics. Several strong changes in absorption are observed in the 59-72\,eV range, appearing to persist only for the duration of the temporal overlap between the XUV and NIR pulses (6-7 fs). Fig. \ref{Stabi}(c) (inverted) shows the result of fitting an exponentially modified Gaussian function to the strong negative signal at 61.7\,eV (coincident with \2). In the fit, the full width at half-maximum (FWHM) of the Gaussian (representative of the instrument response function) is fixed to 5.9\,fs, obtained from measurement of the FWHM of the decay of the He 2\s2\p\ autoionizing state during the same experiment, as shown in Fig. \ref{He} (Appendix \ref{AppA}). The other fitting parameters were allowed to freely vary, yielding a coherence lifetime of 2.4\,\textpm\,0.4\,fs. This coherence lifetime describes the dephasing of the polarization of the XUV-excited dipoles. The ability of this fit to describe the true coherence lifetime, however, is limited by pre- and postpulse structures in the NIR beam. A prepulse is observed at -9\,fs, along with a low-amplitude tail extending out to 25\,fs. Evidence of these features is also present in measurements of He (Fig. \ref{He}), indicating that they are related to the shape of the NIR pulse and not temporal dynamics of the core exciton. A simple Gaussian function is plotted alongside the fit to aid in visualization of the subtle exponential tail of the fit. When comparing this Gaussian to the experimental data, it is evident that, if the coherence of the core exciton persists beyond the approximately 6\,fs temporal overlap of the pump and probe, it is only for a very marginal window of time. Therefore, the 2.4\,\textpm\,0.4\,fs lifetime extracted from the fit serves as an approximate upper limit of the true core exciton coherence lifetime.\par
Despite difficulties in obtaining conclusive temporal information for the \2 feature, the differential absorption data presented is rich in information. A distinct change in absorption is observed at 59.2\,eV, one NIR photon below the 1\s2\s\ dark exciton (60.8\,eV) and where no significant linear absorption is observed in Fig. \ref{Static}. This signal is characterized by clear subcycle oscillations (T\,=\,1.3\,fs), hereinafter referred to as 2$\omega$ oscillations. These are resolved here because of the high degree of temporal stability achieved in the measurements. The negative slope with respect to delay of the fringes indicates that these are the result of couplings to a dark state lying one NIR photon higher in energy \cite{chen_quantum_2013}, supporting that this signal arises as a result of a one-NIR-photon coupling to the 1\s2\s\ dark exciton. A corresponding signal with positively sloped fringes would be expected to appear one NIR photon above the dark state (i.e. ~62.3\,eV). However, this overlaps spectrally with much larger changes in absorption around \2, and because of this it is not resolved here. The remaining transient features also exhibit clear 2$\omega$ oscillations. A myriad of mechanisms may produce these oscillations (including which-way path interferences leading to the same final state \cite{chen_quantum_2013}, the dynamical Franz-Keldysh effect \cite{lucchini_attosecond_2016}, or the oscillating electric field of the NIR pulse); distinguishing between them for such a short-lived signal is difficult and beyond the scope of this study.\par
The origins of the other transient features are less readily apparent. Fig \ref{Stabi}(b) shows the measured differential absorption at representative time delays. The peak near 62.5\,eV closely coincides with \3 and experiences a subtle blueshift near zero delay. This shifting may be indicative of a coupling of \2 to a nearby dark state \cite{moulet_soft_2017, quintero-bermudez_deciphering_2024,geneaux_mgo} or non-resonant Stark effects \cite{quintero-bermudez_deciphering_2024, combescot_semiconductors_1992}. However, because of the small shift and the rapid decay of the transient absorption signals, it is difficult to determine which of these effects is responsible for the observed shift from this measurement alone. The apparent lack of shifting in the remaining spectral features is similarly challenging to interpret fully. To aid in determining the sources of these signals, we examine the effect of altering the relative polarizations of the XUV and NIR beams.\par
\begin{figure}
\includegraphics{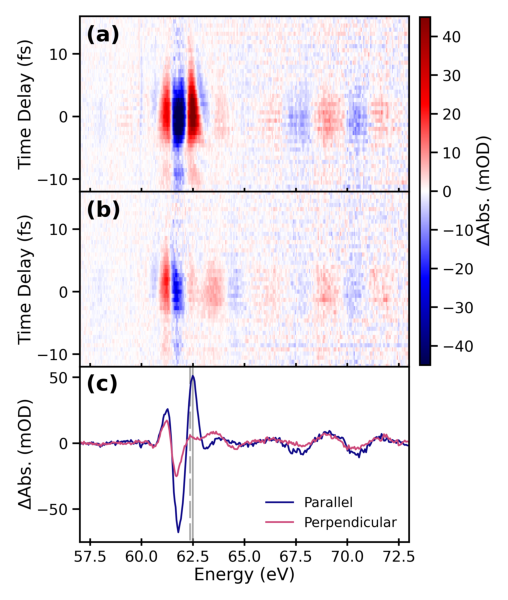}%
\caption{\label{Polar} Polarization dependence of LiF core exciton dynamics for a linear NIR probe polarized parallel (a) or perpendicular (b) relative to a linear XUV pump. (c) Differential absorption for each NIR polarization configuration, taken at $\tau$\,=\,0. The vertical solid (dashed) line shows the energy of the peak of the signal associated with the coupling between \2 and higher lying dark states for parallel (perpendicular) XUV and NIR polarizations, respectively.}
\end{figure}

Fig. \ref{Polar} shows the transient absorption spectrum when using a linear NIR probe polarization that is parallel (a) or perpendicular (b) relative to the polarization of the XUV pump. The NIR probe intensity in this measurement was again set to 6.8\,x\,10\textsuperscript{12}\,W/cm\textsuperscript{2}, and 0.66\,fs time steps were used to sample the delay range of -12\,fs to +33\,fs. To minimize inconsistencies in the experimental conditions, the two NIR polarizations were acquired during the same measurement. The NIR polarization was controlled using an achromatic $\lambda$/2 waveplate ($\lambda$\,=\,500-900\,nm) placed in a motorized rotation mount located in the NIR probe beam path. The waveplate was rotated to 0\textdegree\ and 45\textdegree\ at each time delay point during the measurement to generate the parallel and perpendicular beam alignments, respectively. Temporal stabilization could not be implemented in the polarization dependent measurements. The larger time delay steps and lack of temporal stabilization make the 2$\omega$ oscillations much less apparent here than in Fig. \ref{Stabi}(a). As such, we do not attempt to determine whether the oscillations are absent when using perpendicular beam polarizations.\par 
In the parallel configuration (Fig. \ref{Polar}(a)), the transient features are very similar to Fig. \ref{Stabi}(a). However, when the beams are perpendicularly polarized (Fig. \ref{Polar}(b)), the signal at 62.5\,eV is strongly suppressed. Fig. \ref{Polar}(c) shows the differential absorption of the two NIR polarizations, taken at $\tau$\,=\,0\,fs. There is a clear reduction in intensity of the peak of the positive feature at 62.5\,eV from 51\,\textpm\,5\,mOD in the parallel configuration to 4\,\textpm\,1\,mOD when using crossed polarizations, an approximate 90\% suppression of this peak. This is accompanied by a loss of contrast in the signals appearing at 61.2\,eV and 61.7\,eV. A similar dependence upon the relative polarization of the XUV and NIR beams has been observed for NIR-induced couplings from aligned bright \p- to dark \s-character states in atomic He \cite{reduzzi_polarization_2015}. As in He, this suggests that the 62.5\,eV signal arises as a result of couplings between the 1\s2\p\ bright exciton at 61.7\,eV and a resonant \s-like dark state located near 63\,eV, close to the \3 peak, when the pump and probe polarizations are parallel. When the polarizations of the beams are crossed, this coupling is no longer allowed and thus the large spectral signature is effectively diminished. The suppression of this coupling also accounts for the loss in contrast at 61.7\,eV, as population of \2 is no longer significantly transferred to this dark state. Couplings between \2 and d-like states in LiF are not expected to occur in this energy range. This assertion is based on the energies of the 1\s2\p, 1\s3\p, and 1\s3d transitions of atomic Li\textsuperscript{+}, which occur at 62.22\,eV, 69.65\,eV, and 69.59\,eV, respectively \cite{moore_atomic_1949}. The 2\p\ and 3d states are separated by 7.37\,eV, an energy gap too large to be accessible using the experimental 1.55\,eV NIR pulse. The remaining signals, which appear to be largely unaffected by changes in the probe polarization, are then interpreted to be the result of NIR-induced dynamics that do not involve resonant couplings between states, such as an AC Stark shift of the XUV-excited dipole phases, caused by the ponderomotive shift introduced by the presence of the NIR field \cite{geneaux_mgo}.\par
\subsection{Exciton Computational Modeling}\label{Theory}
Theoretical calculations are employed to gain a deeper understanding of the NIR-induced dynamics taking place. We begin with simulations of a few-level model system to attempt to distinguish differential absorption features arising from couplings between bright and dark states from those caused by the AC Stark effect. Density functional theory (DFT) calculations are then performed to determine the orbital nature of the conduction bands associated with each excitonic feature.\par
\begin{figure*}
\includegraphics{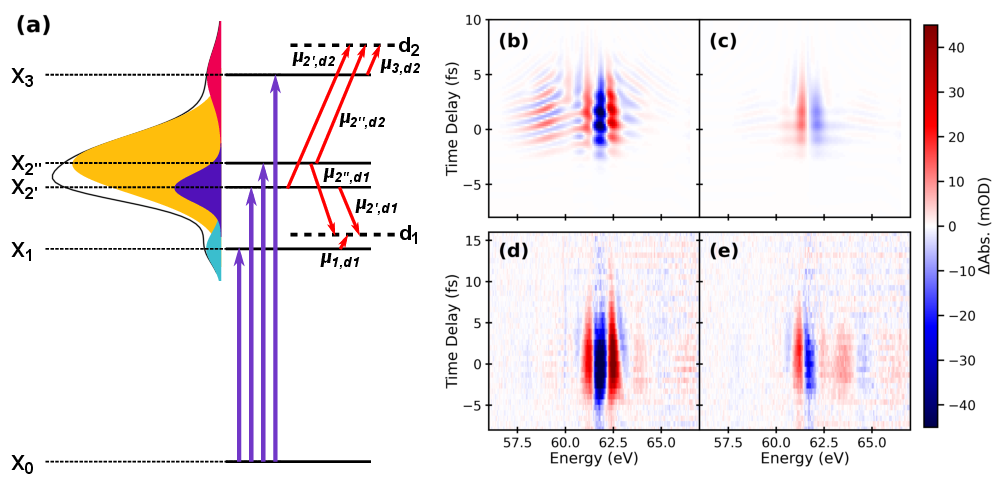}%
\caption{\label{Sim} (a) Energy level diagram of the seven-level system used to model the core exciton dynamics in LiF, induced by XUV pump and NIR probe pulses. The ground state is denoted X\textsubscript{0}, while the four bright exciton states are labeled \1, \2\textsubscript{$'$}, \2\textsubscript{$''$}, and \3. The Gaussian curves identified in the experimental linear absorption profile (black curve) corresponding to these bright states in the model are shown in teal, blue, yellow, and red, respectively. The two dark exciton states are marked d\textsubscript{1} and d\textsubscript{2}. The transition dipole moments between states \textit{p} and \textit{q} are denoted $\mu$\textit{\textsubscript{p,q}}. Computed changes in absorption for (b) the complete model and (c) the model excluding dark state couplings. Experimental transient absorption spectra for the parallel (d) and perpendicular (e) XUV-NIR pulse polarizations. Theoretical results are normalized to the experimental scale.}
\end{figure*}

A few-level model system was used to simulate the experimental transient absorption data in the range of 60-64\,eV, as in Ref. \cite{geneaux_mgo}. To represent the peaks observed in the experimental linear absorption spectrum in this range, the model uses four bright states, directly excited from the ground state X\textsubscript{0}, see Fig. \ref{Sim}(a). In the model, the fourth bright state arises from the asymmetry of the \2 peak in the experimental linear absorption spectrum. DFT calculations of the unbroadened XUV absorption profile presented later in this section indicate that the \2 peak is indeed composed of two underlying sets of peaks, further supporting the inclusion of this additional state. For the model, the experimental \2 peak is thus split into two states, labeled \2\textsubscript{$'$} and \2\textsubscript{$''$}. The energies of the four bright states in the model were determined by fitting the experimental XUV linear absorption profile with four Gaussian functions and extracting the central frequency of each curve, as detailed in Appendix \ref{SimApp}. These are 60.75\,eV, 61.46\,eV, 61.76\,eV, and 62.60\,eV for states \1, \2\textsubscript{$'$}, \2\textsubscript{$''$}, and \3, respectively. Two dark states d\textsubscript{1} and d\textsubscript{2} are also included in the model and coupled to the bright states by the NIR pulse, as described further below. We include these dark states to account for the NIR-induced dynamics observed at 59.25\,eV and 62.5\,eV in the experimental transient absorption results, respectively. d\textsubscript{1} is included to account for the dual bright and dark characteristics of \1 strongly implied by the experimental result. \1 in the experiment is the dipole-forbidden Li 1\s2\s\ dark exciton, appearing in the linear XUV absorption profile because of relaxations in the selection rules caused by lattice distortions \cite{olovsson_near-edge_2009}. Despite appearing in the linear absorption spectrum, transient absorption signals appear in the experiment at 59.25\,eV (one NIR photon below \1), suggesting that one-NIR-photon transitions to the 1\s2\s\ dark state also occur here \cite{wu_theory_2016, chen_quantum_2013}. The dual bright and dark character of the experimental \1 peak is thus treated with two distinct states in the model system. The inclusion of d\textsubscript{1} results in an improvement of the agreement between the simulated and experimental results, as discussed in Appendix \ref{SimApp}. The polarization-dependent measurements suggest that the strong differential absorption signal observed at 62.5\,eV in the parallel XUV and NIR configuration but largely absent when using perpendicular beam polarizations is the result of a coupling between \2 and a higher-lying dark state, near 63\,eV given the experimental NIR pulse energy. This potential dark state is included in the model as d\textsubscript{2}. An energy level diagram illustrating the model system is given in Fig, \ref{Sim}(a).\par
A detailed description of the calculation of the model is provided in Appendix \ref{SimApp}. In brief, the bright states \1-\3 are initially populated from the ground state X\textsubscript{0} via the initial XUV pump pulse, while the dark states d\textsubscript{1} and d\textsubscript{2} are unable to directly interact with the ground state via the XUV pulse. The dark states are instead coupled to the bright exciton states through the NIR probe pulse, as in the experiment. The time-dependent dipole moment for this system obtained by solving the time-dependent Schr{\"o}dinger equation (TDSE) is of the form 
\begin{equation}
\begin{split}
\label{dipole}
D(t,\tau) = -2\mathrm{Re}[\{\sum_{i=1}{c_0^*(t,\tau)}{c_i(t,\tau)}{\mu_{i,0}}e^{i\phi_i(t,\tau)}e^{i{\mathrm{X}_i}t}\}\\
+\{\sum_{j}\sum_{k}{c_{j}^*(t,\tau)}{c_{d_k}(t,\tau)}\mu_{j,d_k}e^{i\phi_j(t,\tau)}e^{i({d_k}-\mathrm{X}_j)t}\}],\\
j = \begin{cases}
1,2',2'' & {k=1}\\
2',2'',3 & {k=2}
\end{cases} \ .
\end{split}
\end{equation}
Here, the index \textit{i} describes the bright exciton states and consists of the values 1, 2$'$, 2$''$, and 3, while the index \textit{k} is associated with the two dark exciton states and includes values 1 and 2. Subscript 0 refers to the ground state. The values of index \textit{j} (referring to conditional sets of the bright exciton states) depend on the value of \textit{k}. When \textit{k}\,=\,1, the index \textit{j} includes 1, 2$'$, and 2$''$; and for \textit{k}\,=\,2, the index \textit{j} takes on the values 2$'$, 2$''$, and 3. This is to account for energetic constraints imposed by the NIR pulse energy: \3 is at too high of an energy to interact with d\textsubscript{1}, and \1 is at too low of an energy to reach d\textsubscript{2} via one NIR photon. The transition dipole moment between states \textit{p} and \textit{q} is denoted $\mu$\textit{\textsubscript{p,q}}, and \textit{c\textsubscript{p}(t,$\tau$)} represents the time- and delay-dependent amplitude of state \textit{p} obtained by integration of the interaction Hamiltonian \cite{wu_theory_2016}. These amplitudes include resonant coupling effects driven by the two-color field. $\tau$ refers to the delay between the XUV and NIR pulses, where a positive value of $\tau$ means the XUV laser pulse arrives before the NIR pulse, as in the experimental results. The phase factor $\phi$\textit{\textsubscript{i}(t,$\tau$)} is incorporated into the dipole moment to account for dynamic phase effects that extend beyond the traditional few-level TDSE model, including non-resonant couplings between electronic states and electron-phonon coupling dynamics. At the phenomenological level, these are defined as
\begin{equation}
    \label{Phases}
    \phi_i(t,\tau) = i\Gamma_{1s}t+\phi_L(t,\tau)+\phi_{ph,i}(t)
\end{equation}
where $\Gamma$\textsubscript{1\s}\,=\,40\,meV is the Li 1\s\
 core-hole Auger-Meitner decay rate \cite{citrin_many-body_1977}; $\phi$\textit{\textsubscript{L}(t,$\tau$)} is the AC Stark phase proportional to the ponderomotive shift imposed by the oscillating NIR field on the electrons; and $\phi$\textit{\textsubscript{ph,i}(t)} describes the coupling between the \textit{i}\textsuperscript{th} state and X-point longitudinal optical phonons \cite{mahan_emission_1977}. Lastly, the attosecond transient absorption spectrogram is calculated as 
 \begin{equation}
 \label{AttoSpec}
 S(\omega,t) = -\omega\mathrm{Im}[E_{\mathrm{XUV}}^*(\omega)\widetilde{d}(\omega,\tau)],
 \end{equation}
where $\widetilde{d}(\omega,\tau)$ and $E_{\mathrm{XUV}}^*(\omega)$ are the Fourier transforms of the dipole moment (Eq. \ref{dipole}) and XUV electric field, respectively \cite{wu_theory_2016}. In the calculation, the energies of d\textsubscript{1} and d\textsubscript{2} are allowed to freely vary to obtain optimum agreement with the experimental transient absorption results. Further details of the calculation and the parameters used are available in Appendix \ref{SimApp}.\par
The results of the model, presented in Fig. \ref{Sim}(b) and \ref{Sim}(c), demonstrate good agreement with the results of the experimental polarization-dependent measurements (shown here again in Fig. \ref{Sim}(d) and \ref{Sim}(e) for comparison). The full simulation (Fig. \ref{Sim}(b)) is able to closely reproduce the characteristic differential absorption signals between 61 and 63\,eV observed in the experimental results using parallel XUV and NIR polarizations (Fig. \ref{Sim}(d)) by allowing for strong couplings between \2 and d\textsubscript{2}. The dark state signal observed in the experimental data at 59.2\,eV is also present in the simulated data. This signal is attributed to one-NIR-photon couplings to d\textsubscript{1} here, consistent with the description above of one-NIR-photon transitions to the 1\s2\s\ dark state resulting in a light-induced signal at this energy in the experiment. In the model, the energy of d\textsubscript{1} is found to be 60.83\,eV, 0.08\,eV above the central frequency of \1. We posit that the very slight energetic separation between d\textsubscript{1} and \1 in the model is related to minor differences in energy between the transition allowed by lattice distortions and the "true" 1\s2\s\ dark state. d\textsubscript{2} is found to lie at 63.23\,eV, one NIR photon above \2\textsubscript{$''$}. Signals arising as a result of two-photon transitions to \2\textsubscript{$''$} are also observed at 59\,eV in the simulation, but are not resolved in the experimental results. Fig. \ref{Sim}(c) gives the result of the model when dark state coupling is excluded from the dipole moment calculation, leaving the AC Stark phase as the only delay-dependent dynamic phase remaining in the calculation. This configuration closely resembles the results of measurements utilizing perpendicular XUV and NIR polarizations (Fig. \ref{Sim}(e)). The positive signal at 62.5\,eV is notably absent here, indicating that this transient feature is largely the result of the aforementioned coupling between \2 and a higher-lying dark state, while the remaining signals are produced through a Stark shift of the \2 bright exciton, consistent with interpretations of the experimental results.\par
\begin{figure}
\includegraphics{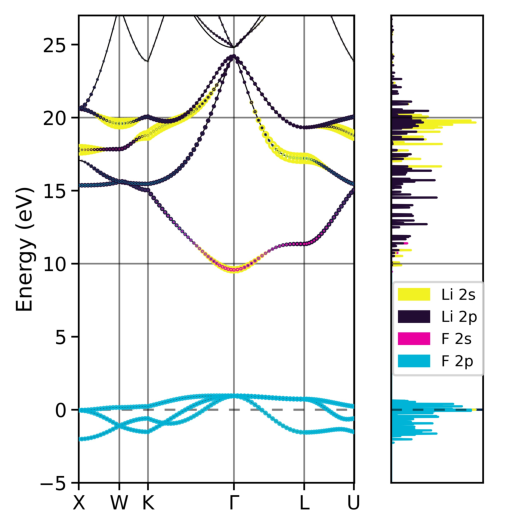}%
\caption{\label{Bands} Calculated band structure and projected density of states of LiF, describing the orbital character of the valence and conduction bands.}
\end{figure}
\begin{figure}
\includegraphics{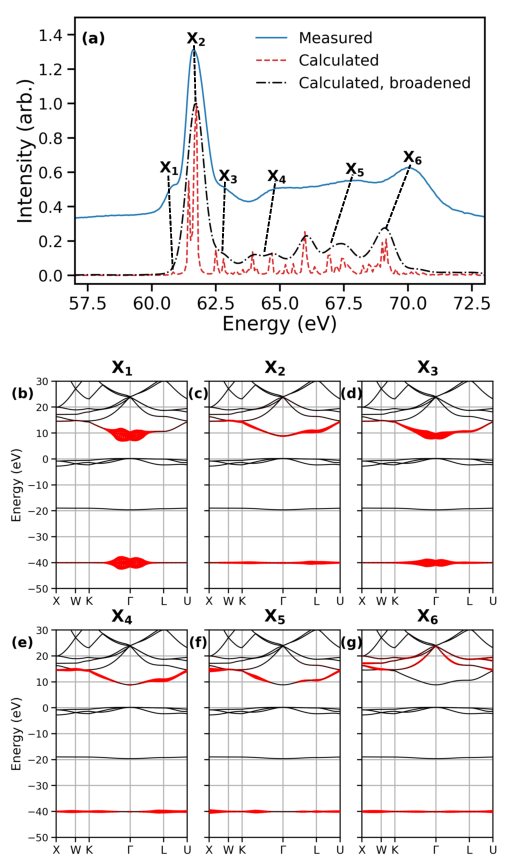}%
\caption{\label{Weights} Computational modeling of LiF core excitons. (a) Calculated (red dashed curve) and experimental (solid blue curve) linear absorption spectrum. The calculated spectrum convoluted with a $\sigma$\,=\,300\,meV Gaussian is given by the black dash-dotted curve, demonstrating close agreement with the experimental curve. Black dashed lines connect \1-\6 to representative excitons identified in the simulated spectrum. A 14\,eV scissor correction \cite{wang_real_2019} has been applied to the calculated spectra in (a) to compensate for underestimation of the bandgap. (b)-(g) Excitonic weights (red) projected onto the LiF band structure for \1-\6, sequentially. In each, heavier red curves indicate regions with greater contributions to the wavefunction of that particular exciton. No bandgap corrections have been applied in (b)-(g).}
\end{figure}

To understand the nature of the dark state coupled to \2, the projected density of states (pDOS) and weights of the excitonic wavefunctions are calculated in a procedure that closely follows that of Ref. \cite{quintero-bermudez_deciphering_2024}. Comparison of the excitonic weights (Fig. \ref{Weights}) and pDOS calculations (Fig. \ref{Bands}) enables analysis of the orbital composition of core excitonic features present in the XUV absorption profile. The band structure and pDOS were computed using the Quantum Espresso software package \cite{giannozzi_quantum_2009,giannozzi_advanced_2017}. The calculated band structure near and above the band gap is shown in Fig. \ref{Bands}
and apart from underestimation of the band gap is consistent with previous calculations \cite{wang_quasiparticle_2003,sommer_optical_2012}. Along with the valence and conduction bands, the projection includes the Li 1\s\ and F 2\s\ core levels. To aid viewing the details of the CB, these core levels are omitted in Fig. \ref{Bands}. The complete band structure including the core levels is available in Fig. \ref{FullBand} (Appendix \ref{DFT Methods}). The valence band consists primarily of F 2\p\ orbitals. The CB edge is dominated by Li 2\p\ orbitals, apart from a large density of Li 2\s\ character near the $\Gamma$-point. The projection also places small F 2\s\ density near the CB minimum at $\Gamma$. Other calculations of the pDOS of LiF include F \s-like contributions near the band edge, but do not identify the principal quantum number \textit{n} value associated with these bands \cite{pascal_finite_2014,rajput_two-dimensional_2022}. It is unlikely that the filled F 2\s\ core level would contribute to the conduction band, so the presence of F 2\s\ orbitals in the CB here is attributed to a mathematical artifact of the projection operation and considered nonphysical. At higher energies, the CB is composed primarily of Li 2\p\ orbitals, with small regions of localized Li 2\s\ character. The projection operators in the pDOS calculation include only the Li 1\s, 2\s, and 2\p\ orbitals, and the F 2\s\ and 2\p\ orbitals. As a result, the \textit{n}\,=\,3 orbitals of each species are not present in the pDOS. As previously stated, based on the atomic lines for Li\textsuperscript{+}, the 1\s3\p\ and 1\s3d transitions are expected to lie high in the conduction band, several eV above the 1\s2\p\ bright exciton near the CB edge.\par
Excitonic weight contributions to the linear absorption spectrum were determined using spin-singlet BSE calculations via the Exciting computational package \cite{gulans_exciting_2014,draxl_exciting_2017,vorwerk_addressing_2017}. These are shown in Fig. \ref{Weights}.
The computed linear absorption spectrum is shown in Fig. \ref{Weights}(a) alongside the experimental XUV absorption spectrum, showing that the two are in excellent agreement. Labels \1-\6 are the same as in Fig. \ref{Static} and connect core excitonic features identified in the calculation to corresponding structures observed in the measured linear absorption spectrum. A constant 14 eV scissor correction \cite{wang_real_2019} was applied to the calculated spectrum to compensate for energetic differences due to underestimation of the band gap and Li 1\s\ core level energies. The correction was optimized to obtain agreement between the experimental and calculated \2 peaks, leading to slight deviations in the energies of \1 and \3-\6. The calculated spectrum is convolved with a Gaussian ($\sigma$\,=\,300\,meV) to better reflect the inhomogeneous broadening of the experimental system. The unbroadened spectrum is given as well. It is evident from the calculation that the LiF XUV absorption profile is composed of several excitonic features. The peaks in the calculated spectrum for \2-\6 consist of sets of multiple excitonic features. The orbital compositions of the excitons within each set are found to be very similar, so to facilitate analysis, representative excitons of each set are given and described below. Figs. \ref{Weights}(b)-\ref{Weights}(g) show the excitonic weights of \1-\6 plotted on the calculated band structure, with heavier red curves indicating a larger contribution of that band to the excitonic wavefunction.\par
Considering both the pDOS and BSE calculations, we find that the nature of \1 and \2 are in good agreement with previous studies \cite{pantelides_electronic_1975,haensel_measurement_1968,gudat_core_1974,fields_electronic_1977,rottke_probing_2022}. \1 is found to be strongly localized around the $\Gamma$-point and consists primarily of Li 2\s\ orbitals, as expected for the 1\s2\s\ exciton. \2 corresponds to conduction bands composed primarily of Li 2\p\ orbitals and is delocalized across the Brillouin zone. \3 corresponds to the higher-energy shoulder of the main bright exciton, which has not been investigated in detail previously. This peak is composed of multiple excitonic states in the range of 62.5-63 eV, and like \1, these are characterized by significant contributions of Li 2\s-like bands, indicating that this feature is mainly formed by dark excitonic states. However, the states forming the \3 peak are slightly more delocalized and exhibit greater Li 2\p\ character than \1, explaining why \3 is clearly visible in the calculated linear absorption spectrum (Fig. \ref{Weights}(a)) while \1 is not. This distinction is less prevalent in the experimental XUV absorption profile (Fig. \ref{Static}) because of symmetry-breaking lattice distortions \cite{olovsson_near-edge_2009} that relax the selection rules regarding the Li 1s to 2s transition. It is of note that the \s-like excitonic states forming \3 extend to energies close to the energy of d\textsubscript{2} determined in the few-level model simulation. Significant excitonic contributions to the linear absorption profile are also identified between 64\,eV and 69\,eV, as was initially postulated by Pantelides \cite{pantelides_electronic_1975}. Excitons in this region are dominated by Li 2\p\ orbitals and are delocalized across the CB edge, as represented by \4 and \5. Finally, core exciton-like states lying high in the CB are shown to contribute to the formation of the \6 peak. The pDOS calculations presented in Fig. \ref{Bands} suggest that this feature occupies bands of primarily Li 2\s\ and 2\p\ orbital character. It is noted again that the pseudopotentials used in the pDOS and exciton weight calculations include projections for only the \textit{n}\,=\,1 and \textit{n}\,=\,2 atomic levels of Li and F. The results of these calculations provide support for the interpretation of the polarization dependent results presented in Section \ref{Experiment}.\par
\section{Discussion}
\subsection{Polarization-Dependent Core Exciton Dynamics}
The LiF linear XUV absorption profile is characterized by six distinctive features: \1, the 1\s2\s\ dark exciton, observed despite being dipole-forbidden because of symmetry-breaking lattice distortions \cite{olovsson_near-edge_2009}; \2, the 1\s2\p\ bright exciton; \3, which has not been previously discussed in the literature, but which we find to be formed by excitons of primarily Li 1\s2\s\ character; \4 and \5, shown to contain Li 1\s2\p\ core excitonic transitions; and \6, which has been described as an electronic polaron \cite{kunz_role_1972}. The original description of \6 was "a band electron dressed by a cloud of virtual excitons (electronic polaron)", and this polaron complex was said to be bound to the 1\s\ core hole \cite{kunz_absorption_1973}. However, a recent four-wave mixing study found insufficient experimental evidence for the colocalization required for such a description to be feasible \cite{rottke_probing_2022}. As an alternative explanation for this feature, we consider the atomic Li\textsuperscript{+} 1\s3\p\ transition at 69.37\,eV \cite{moore_atomic_1949}, which is in close proximity energetically to \6. The excitonic weight calculation finds that the \6 electron lies in bands largely composed of Li 2\s\ and 2\p\ orbitals. The pseudopotentials used in the pDOS calculation do not account for the \textit{n}\,=\,3 orbitals of either Li or F. We speculate that bands in this region also exhibit significant Li 3\p\ character, based on the atomic 1\s3\p\ transition energy.\par
The transient absorption results are shown to have a clear dependence upon the polarization of the NIR probe relative to the XUV pump, evident in the strong suppression of the signal at 62.5\,eV. We begin the discussion of this effect by first describing similar results in an earlier study of gaseous He. In He, it was shown that using perpendicular XUV and NIR polarizations results in suppression of couplings from aligned \p-orbitals to \s-orbitals, while couplings to d-orbitals are mostly unaffected \cite{reduzzi_polarization_2015}. This is understood by considering a scheme in which the NIR polarization is always aligned along the \textit{z}-axis, forming the quantization axis, while the XUV polarization is allowed to vary. When the polarizations of the two beams are parallel along the \textit{z}-axis, the XUV pulse excites from the He 1\s\ core level (\textit{l}\,=\,0, \textit{m}\,=\,0) into states in the \textit{n}p Rydberg series with \textit{m}\,=\,0. The NIR probe is then able to couple from the XUV-excited \textit{l}\,=\,1, \textit{m}\,=\,0 states to the \textit{m}\,=\,0 component of nearby \s- and d-orbital dark states. When the beam polarizations are perpendicular (XUV along the \textit{x}-axis), the excited \textit{n}p Rydberg states prepared by the XUV pulse now have \textit{m}\,=\,\textpm1. The NIR is still polarized along \textit{z} and thus cannot change the value of \textit{m}, meaning that NIR-induced couplings to \s-orbitals are now forbidden, and differential absorption signals associated with these transitions are no longer observed. Couplings to \textit{d}-aligned dark states are unaffected by the change in polarization, as there are \textit{m}\,=\,\textpm1 components available.\par
A similar excitation scheme is present in LiF, where, in the simplest description, the XUV field excites electrons from the Li 1\s\ core level into bands that are primarily Li 2\p-like, forming \2. A key distinction is that, in LiF, there are no energetically accessible \textit{d}-orbitals for the NIR to couple with \cite{moore_atomic_1949}, and instead only couplings from \2 to \s-like states may occur. When using perpendicular XUV and NIR polarizations, the strong differential absorption observed at 62.5\,eV is reduced by 90\%, indicating that a coupling between \2 (the 1\s2\p\ bright exciton at 61.7\,eV) and an \s-like dark state one NIR photon higher in energy has been (largely) suppressed. Exciton weight calculations reveal that \3, located near 63.2\,eV, is composed primarily of excitonic states across the 62.5-63\,eV range associated with Li 2\s-like bands, verifying that dipole-allowed couplings between \2 and \3 are indeed possible. Simulations of a model few-level system provide further support for the strong resonant coupling between \2 and \3. Results of the simulation when using the full model closely resemble the experimental parallel polarization results. When the coupling to the higher energy dark state is removed from the model, the strong signal at 62.5\,eV associated with the coupling from \2 to \3 is largely absent, and the overall results are in good agreement with the perpendicularly polarized experimental data. Unlike in He, there is still a small ($\sim$4\,mOD) signal observed at 62.5\,eV when using perpendicular XUV and NIR polarizations. Exciton weight and pDOS calculations indicate that \2 and \3 are not purely \p- and \s-like, respectively, which complicates the coupling scheme in LiF compared to the atomic couplings in He. The bands associated with \2 show weak Li 2\s-like contributions, and the CB minimum occupied by \3 is found to contain small Li 2\p-like character. Couplings between these comparatively weaker contributions to \2 and \3 would then be polarization-invariant, accounting for the weak differential absorption signal observed around 62.5\,eV when using perpendicular beam polarizations. It is also possible that other underlying NIR-induced effects not sensitive to changes in relative polarization may produce this small signal, such as broadening of the \2 peak \cite{zurch_direct_2017}. Similarly, we speculate that the lattice distortions responsible for the observation of the 1\s2\s\ dark state in linear absorption measurements may also relax the selection rules of the coupling between \2 and \3.\par
\subsection{Core Exciton Decoherence}
The Li 1\s2\p\ core exciton lifetime is found to extend only slightly beyond the duration of the overlap of the pump and probe beams, meaning the 2.4\,\textpm\,0.4\,fs signal duration obtained from the fit can be treated only as an experimental upper limit of the coherence lifetime. As a result absolute conclusions regarding the mechanism responsible for the rapid transient absorption signal decay cannot be drawn from the present study. We draw on existing core exciton measurements to facilitate a preliminary discussion of the dynamics observed. Historically, discussions have considered either Auger-Meitner decay of the core hole \cite{citrin_interatomic_1973} or a phonon-mediated dephasing of the exciton \cite{mahan_emission_1977} to be responsible for the broad absorption linewidths noted for core excitonic features in ionic insulators. For Auger-Meitner decay, the exciton would be destroyed by the filling of the core hole. This is in contrast to the phonon model, in which couplings between the exciton and phonons lead to a rapid dephasing of the polarization of the excited core excitons.\par
Descriptions of the phonon dephasing mechanism have become ambiguous through successive interpretations. To provide clarity, we present a brief overview of the development of the phonon coupling model. Early discussions of the ties between phonons and core excited states in the 1970s focused on identifying the origin of the excess broadening observed in X-ray photoemission lines in solids, as lifetime broadening and charging effects were found to be insufficient to produce the observed profiles. Analogous to excitations involving color centers in alkali halides, it was determined that formation of the core hole is accompanied by simultaneous production of a large number of phonons \cite{matthew_breadths_1974,citrin_phonon_1974,mahan_emission_1977,citrin_many-body_1977,matthew_temperature_1970,almbladh_effects_1977,mahan_photoemission_1980}. These phonons arise from the structural rearrangement of the lattice due to Coulombic attraction between the core hole and surrounding electrons. The coupling between the core hole and these phonons was found to broaden both the absorption and emission spectra of the system. This coupling is typically described as a phase factor $\phi$\textit{\textsubscript{ph,i}(t)} in the absorption, given by Mahan \cite{mahan_emission_1977} as
\begin{equation}\label{Mahan}
\begin{split}
    \phi_{ph,i}(t)=i\frac{M_i^2}{\omega_i^2}[(2N+1)(1-\cos{\omega_it})-i(\omega_it-\sin{\omega_it})].
\end{split}
\end{equation}
Here, \textit{M\textsubscript{i}} and $\omega$\textsubscript{\textit{i}} are the coupling strengths between the phonon and \textit{i}\textsuperscript{th} excited electronic dipole and the phonon frequency, respectively, and \textit{N} is the thermal phonon population. It then follows that core excitons in ionic insulators, and particularly in the alkali halides, are expected to exhibit strong phonon coupling. In these materials, the core hole resides on the cation and is necessarily poorly screened from the surrounding electron-rich anions due to the cation's reduced electron density, suggesting a large number of phonons will be produced in response to the formation of the core hole. This strong phonon coupling would then also account for the broad absorption peaks associated with core excitons \cite{rubloff_far-ultraviolet_1972,pantelides_electronic_1975,haensel_measurement_1968,gudat_core_1974,pantelides_new_1974,kowalczyk_x-ray_1974,stott_core_1984}. More recently, the phonon coupling model has been extended to time-resolved core exciton studies to account for the short-lived signals observed \cite{moulet_soft_2017,geneaux_mgo, quintero-bermudez_deciphering_2024}. In these studies, it was determined that the phonon coupling not only broadens the linear absorption profile of the exciton as previously noted, but also leads to a rapid dephasing of the XUV-excited dipoles. The resulting incoherent ensemble of core excitons is then no longer able to be detected using transient absorption or reflectivity methods.\par
To consider how Auger-Meitner decay and phonon-mediated dephasing relate to the results obtained in LiF, we draw analogies to results obtained in MgO, which exhibits the same halite crystal structure as LiF. The core exciton coherence lifetimes in MgO were found to be 1.6\,\textpm\,0.5 and 2.3\,\textpm\,0.2\,fs \cite{geneaux_mgo}, close to the LiF estimate of 2.4\,\textpm\,0.4\,fs. The Auger-Meitner linewidths of the metallic Li 1\s\ and Mg 2\p\ core holes are 40\,meV (8.2\,fs) and 30\,meV (10.97\,fs), respectively \cite{citrin_many-body_1977}. In both cases, the core hole lifetime obtained solely from the Auger-Meitner linewidths far exceeds the measured core exciton lifetimes. While Auger-Meitner decay processes have been noted for LiF \cite{citrin_interatomic_1973,gallon_low_1970,matthew_transition_1975,hotokka_auger_1984}, this is less likely to play a major role due the rapid decay of the LiF core exciton transient absorption signals observed. Instead, Auger-Meitner decay serves as a separate, slower method by which the core hole (which may persist beyond the decoherence of the exciton) may decay. In MgO, along with other time-resolved core exciton studies of SiO\textsubscript{2} \cite{moulet_soft_2017} and CaF\textsubscript{2} \cite{quintero-bermudez_deciphering_2024}, it was also determined that the rapid decay of the core exciton signals was the result of strong couplings to phonons, as described by Mahan \cite{mahan_emission_1977}. Core excitons in halite crystals like LiF and MgO have been determined to couple most strongly to X-point longitudinal optical phonons $\omega$\textsubscript{LO} \cite{mahan_photoemission_1980}. The phonon energies of LiF ($\omega$\textsubscript{LO}\,=\,80\,meV \cite{dolling_lattice_1968,willett-gies_two-phonon_2015}) and MgO ($\omega$\textsubscript{LO}\,=\,60\,meV \cite{singh_crystal_1972}) are of similar magnitude, as are the lifetimes of their core exciton signals. These similarities suggest that a phonon-mediated dephasing mechanism may also drive the rapid coherence decay of the LiF core exciton, as suggested in other studies. A more rigorous understanding of the decoherence would require further measurements with even shorter temporal resolution, as in an attosecond pump-attosecond probe scheme, as well as temperature-dependent measurements to further elucidate the possible role of the lattice's phonon population. The thermal phonon population \textit{N} is inherently temperature dependent, indicating the phonon coupling phase in Eq.\,\ref{Mahan} is also sensitive to changes in temperature. Intuitively, if the short-lived transient absorption signal lifetimes observed in LiF are due to a phonon-mediated decoherence, it is expected that the dephasing should occur more rapidly at higher temperatures (larger \textit{N}) and more slowly at lower temperatures (smaller \textit{N}). Similarly, theoretical investigations may provide greater insight into the source of the rapid signal decay. 
\section{Conclusion}
XUV transient absorption spectroscopy was used to investigate core exciton dynamics of LiF at the Li K edge. Several core excitons excited from the Li 1\s\ core level were identified, spanning the linear XUV absorption profile between 60 and 72\,eV. The rapid decay of the core exciton coherence is preliminarily concluded to be the result of a phonon-mediated dephasing mechanism, though further investigations are needed to validate this determination. The orbital character of a dark state resonant with the 1\s2\p\ bright exciton was experimentally observed by controlling the relative polarization of the XUV pump and NIR probe fields. When the polarization of the two beams was crossed, the differential absorption signal associated with this coupling was reduced by 90\%, indicating coupling between the 2\p-aligned bright state and an \s-orbital like dark state was efficiently suppressed. Theoretical calculations verified the presence of previously unidentified Li 2\s-like states within 1 NIR photon resonance of the Li 1\s2\p\ bright exciton. These results indicate that laser polarization serves as a powerful experimental tool for probing the orbital alignment of absorption features in the condensed phase, expanding upon previous investigations conducted in atomic systems.\par
\begin{acknowledgments}
Experimental investigations were supported by the Air Force Office of Scientific Research (AFOSR) Grant Nos. FA9550-24-1-0184, FA9550-19-1-0314, and FA9550-20-1-0334. LD acknowledges the European Union's Horizon research and innovation programme under the Marie Sk\l{}odowska-Curie grant agreement No. 101066334--SR-XTRS-2DLayMat. DFT calculations were performed through the UC Berkeley College of Chemistry Molecular Graphics and Computational Facility, supported by NIH S10OD034382. Theoretical work at LSU was supported by the U.S. Department of Energy, Office of Science, Basic Energy Sciences under Contract No. DE-SC0010431.
\end{acknowledgments}

\appendix

\section{Experimental Methods}\label{AppA}

\begin{figure}
\includegraphics{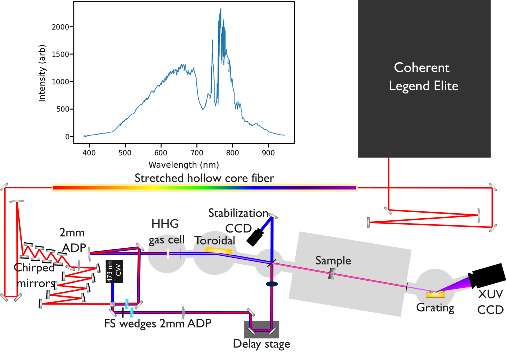}%
\caption{\label{Setup} Schematic representation of the experimental setup. A typical spectrum measured after spectral broadening in the hollow core fiber is also included. The $\lambda$/2 waveplate placed in the NIR arm is used only in the polarization dependence measurements. The 473\,nm CW beam is used only in the measurement with temporal stabilization.}
\end{figure}
The output of a Ti:Sapphire laser (Coherent Legend Elite, 5\,mJ pulse energy, 25\,fs pulse duration FWHM, 1\,kHz repetition rate, 795\,nm central wavelength) is used to produce the XUV pump and NIR probe pulses used in these experiments. The laser output is focused into a 2\,m long stretched hollow core fiber (inner diameter 530\,{\textmu}m) filled with a He pressure gradient (0.5\,bar at optical entrance, 2.5\,bar at optical exit). This results in an approximately 2\,mJ pulse spanning 450-950\,nm (Fig. \ref{Setup}).\par
The dispersion introduced in the fiber is initially compensated for by passing through 8 pairs of broadband double angle chirped mirrors (Ultrafast Innovations PC70, PC1332), along with a 2\,mm thickness of ammonium dihydrogen phosphate (ADP) to correct for third order dispersion \cite{timmers_generating_2017}. The beam is then passed through a mechanical chopper, reducing the repetition rate to 500\,Hz to reduce sample degradation caused by heating. The pump and probe arms are produced by an 80:20 beamsplitter, respectively. Each arm has a pair of anti-reflection coated fused silica (FS) wedges used to compensate for second order dispersion, resulting in $\sim$4\,fs NIR pulses.\par
The 1.6\,mJ of NIR transmitted by the beamsplitter is focused by a 0.5\,m focal length mirror into a vacuum chamber housing a 4\,mm long gas cell with $\sim$500\,{\textmu}m aperatures with $\sim$31\,Torr Ar gas pressure for high harmonic generation. The resulting XUV pump beam continuously spans photon energies from $\sim$20-72 eV. The remaining NIR driving field is filtered by a 150\,nm thick Al foil (Lebow Company), and the transmitted XUV is focused by a gold-coated toroidal mirror onto the sample. The 0.4\,mJ of NIR reflected by the beamsplitter is  used as the probe pulse, after further attenuation by an iris. An additional 2\,mm thickness of ADP is added to the NIR probe arm to further compensate for third order dispersion in the beam. The time delay between the XUV and NIR pulses is controlled by a two-mirror retroreflector mounted on a piezoelectric stage (Physik Instrumente P-620.1CD) located in the probe arm. The NIR probe is then focused by a 1\,m focal length mirror and recombined collinearly with the XUV pump by reflection from an annular mirror. After transmission through the sample, the residual NIR probe is blocked by a 150\,nm Al foil, and the XUV transmitted through the filter is spectrally dispersed by a gold-coated grating (Hitachi 001-0640) onto an XUV charge-coupled device camera (Princeton Instruments PIXIS-XO 400B).\par
\begin{figure}
\includegraphics{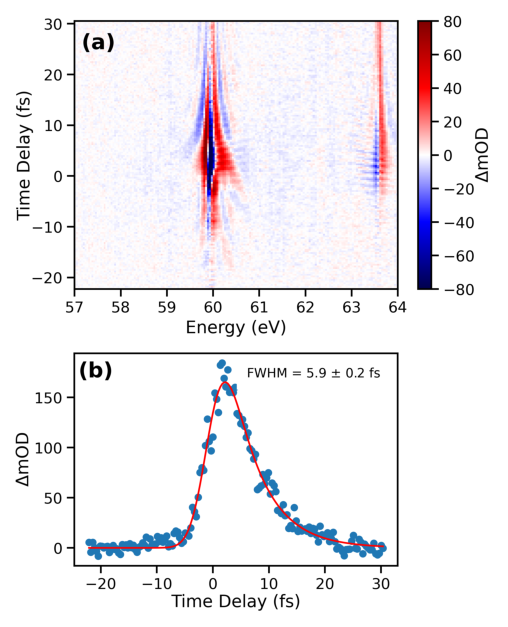}%
\caption{\label{He} (a) Decay of the 2\s2\p\ autoionizing state in He, used to optimize the spatial overlap of the XUV and NIR beams, as well as determine the duration of the temporal overlap of the two beams. The overlap duration is extracted from the full width at half-maximum of the exponentially modified Gaussian function (red curve) used to fit the decay of the state (blue dots), as shown in (b). The FWHM of the Gaussian is in the range of approximately 5-8 fs for the experiments presented here.}
\end{figure}

Temporal and spatial overlap is determined by measuring the decay of the doubly-excited 2\s2\p\ autoionizing state in He \cite{gilbertson_monitoring_2010,kaldun_observing_2016}, shown in Fig. \ref{He}. 
A prepulse is observed near -10\,fs, as also noted in the LiF measurements. The delay-dependent signal is fit with an exponentially modified Gaussian to determine the position of time zero and the duration of temporal overlap. A weak tail found to persist until $\sim$20\,fs after t\textsubscript{0} in LiF is also observed in the lineout in Fig. \ref{He}(b), suggesting that this feature in the LiF data is related to the NIR pulse shape, rather than any core exciton-related dynamics. For measurements where delay stabilization is not implemented, He measurements collected after each scan cycle are used to correct for drifts in temporal overlap. The full width at half-maximum (FWHM) of the Gaussian component of the fit is used to determine the cross correlation of the pump and probe beams, which is $\sim$5-8\,fs for the results presented here. The samples measured are 20\,nm thick polycrystalline LiF thin films (Lebow Company) evaporated onto 30\,nm thick Si\textsubscript{3}N\textsubscript{4} membranes (Norcada NX5050X). The samples and gas cell are mounted on an \textit{xy} translation stage (Physik Instrumente).
\par

\section{Computational Methods}

\subsection{Few-Level Model Simulation}\label{SimApp}
As discussed in the main text, a two-step fitting procedure was used to determine the unknown parameters in the theoretical model. First the parameters of the bright states are extracted by fitting the experimental XUV-only linear absorption spectrum in the 60-64 eV energy range. Next, the energies of two dark states, their couplings to the bright exciton states, and the Stark shift constant are determined by fitting the experimental transient absorption spectra.\par
We first discuss the fitting procedure used to determine the bright state parameters using only the linear XUV absorption data, with particular focus on the energies of the bright states and the transition dipole moments of the ground-to-excited bright state transitions. The dipole moment for this system is given as
\begin{equation}
\label{XUVdipole}
d(t,\tau) = -2\mathrm{Re}[\sum_{i=1}{c_0^*(t,\tau)}{c_i(t,\tau)}{\mu_{i,0}}e^{i\phi_i(t,\tau)}e^{iX_it}].
\end{equation}
Here, the index 0 refers to the ground state, and index \textit{i} includes the values 1, 2$'$, 2$''$, and 3, corresponding to the four directly-excited bright exciton states considered in the model. X\textsubscript{\textit{i}} refers to the XUV-excited bright exciton states. X\textsubscript{2$'$} is introduced to the model system to account for asymmetry of the \2 peak in the experimental linear absorption profile. The transition dipole moment between the ground and \textit{i}\textsuperscript{th} excited state is denoted by $\mu$\textsubscript{0,\textit{i}}, and \textit{c\textsubscript{i}(t,$\tau$)} represents the time- and delay-dependent amplitude of state \textit{i} obtained from integration of the interaction Hamiltonian \cite{wu_theory_2016}. These include resonant couplings between electronic states caused by the two-color field. \textit{$\phi$\textsubscript{i}(t,$\tau$)} are the dynamic phases not accounted for by the time-dependent Schr{\"o}dinger equation, including non-resonant electronic and electron-phonon couplings. At the phenomenological level, these phases are defined as
\begin{equation}
    \label{phases}
    \phi_i (t,\tau)= i\Gamma_{1s}t+\phi_L(t,\tau)+\phi_{ph,i}(t). 
\end{equation}
$\Gamma$\textsubscript{1\s}\,=\,40\,meV is the Auger-Meitner decay rate for the metallic Li 1\s\ core level \cite{citrin_many-body_1977}. $\phi$\textit{\textsubscript{L}(t,$\tau$)} is the AC Stark phase, proportional to the ponderomotive shift \textit{U\textsubscript{p}} on the electrons imposed by the oscillating NIR field, defined as
\begin{equation}
    \label{stark}   
    \phi_L (t,\tau)=-\alpha\int_{0}^{t}U_p (\tau,t')dt'.
\end{equation}
Here, $\alpha$ is the AC Stark phase constant to be determined by fitting. Finally, $\phi$\textit{\textsubscript{ph,i}(t)} is the phase corresponding to the exciton-phonon coupling described by Mahan \cite{mahan_emission_1977} and defined as
\begin{equation}
    \label{Mahanapp}
    \phi_{ph,i}(t)=i\frac{M_i^2}{\omega_{LO}^2}[(2N+1)(1-\cos{\omega_{LO}t})-i(\omega_{LO}t-\sin{\omega_{LO}t})].
\end{equation}
It has been determined that formation of the core hole in ionic insulators is accompanied by the production of a large number of phonons \cite{mahan_emission_1977, matthew_breadths_1974,citrin_phonon_1974,citrin_many-body_1977,mahan_photoemission_1980,matthew_temperature_1970,almbladh_effects_1977}. The effect of this coupling between the core hole and these phonons on absorption and emission of the system is described by the phonon phase factor in Eq. \ref{Mahanapp}. Here, \textit{M\textsubscript{i}} are constants describing the coupling between the bright exciton states X\textsubscript{i} and the phonons, \textit{N} is the thermal phonon population, and $\omega$\textsubscript{LO}\,=\,80\,meV is the Li X-point optical phonon energy \cite{dolling_lattice_1968,willett-gies_two-phonon_2015}.\par
\begin{figure}
\includegraphics{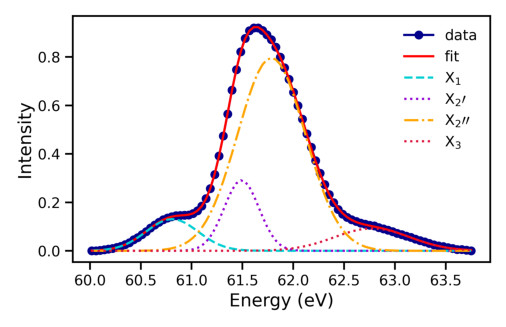}%
\caption{\label{FitStat} The simulated XUV-only absorption spectrum, obtained using the parameters listed in Table \ref{XUVTable}, is given by the solid red curve. The experimental data points are given by the dark blue dots. The four Gaussian curves used to obtain the simulated spectrum are given by the teal dashed curve, purple dotted curve, yellow dash-dotted curve, and red dotted curve, corresponding to \1, \2\textsubscript{$'$}, \2\textsubscript{$''$}, and \3, respectively.}
\end{figure}

 
The attosecond transient absorption spectrogram \textit{S($\omega$,$\tau$)} is calculated using the Fourier transform of the dipole moment and the XUV electric field, as detailed in \cite{wu_theory_2016}. The calculated spectrum is optimized to fit the experimental linear XUV absorption profile between 60 and 64\,eV. We first subtract a linearly increasing background from the experimental absorption spectrum so that the absorption profile goes to zero outside the multi-peak feature spanning 60-64 eV. Four Gaussian peaks are fit to the background-subtracted experimental spectrum (Fig. \ref{FitStat}), obtaining the transition energy, bandwidth, and relative strengths of the four peaks.
Because the dipole moment is represented in the time domain, the Gaussian functions were then Fourier transformed into the time domain, and the amplitude, width, and central frequencies were used as the initial parameters for the XUV-only fitting procedure. The results of this fitting are given in Table \ref{XUVTable} and are in good agreement with the experimental linear XUV absorption data, as shown in Fig. \ref{FitStat}. Note that the overall strength of the calculated response has an arbitrary scaling factor relative to the experimental one. This is represented through the fixed value of the largest coupling matrix element $\mu$\textsubscript{0,2$''$}.\par 
\begin{table}
\caption{\label{XUVTable} Results and confidence intervals (C.I.) of fitting the experimental linear XUV absorption spectrum to determine the energies, transition dipole moments $\mu$\textsubscript{0,\textit{i}}, and phonon coupling constants \textit{M\textsubscript{i}} of \1-\3. The transition dipole moments $\mu$\textsubscript{p,q} are given in atomic units}
\begin{ruledtabular}
\begin{tabular}{ c c c }
Paramter & Value & C.I.\\
\hline
\1 (eV) & 60.75 & 0.023\\
X\textsubscript{2$'$} (eV) & 61.46 & 0.01\\
X\textsubscript{2$''$} (eV) & 61.76 & 0.03\\
\3 (eV) & 62.60 & 0.09\\
$\mu$\textsubscript{0,1} & 0.187 & 0.002\\
$\mu$\textsubscript{0,2$'$} & 0.259 & 0.001\\
$\mu$\textsubscript{0,2$''$} & 0.559 & fixed\\
$\mu$\textsubscript{0,3} & 0.248 & 0.003\\
M\textsubscript{1} (eV) & 0.199 & 0.016\\
M\textsubscript{2$'$} (eV) & 0.166 & 0.017\\
M\textsubscript{2$''$} (eV) & 0.259 & 0.018\\
M\textsubscript{3} (eV) & 0.417 & 0.053\\
\end{tabular}
\end{ruledtabular}
\end{table}
The dark exciton states do not interact by one photon with the ground state: instead, they are coupled to the bright exciton states via the NIR probe pulse. The experimental transient absorption spectra are used to accurately determine the position of the dark states, their coupling to the bright states, and the Stark shift imposed by the presence of the NIR field. In order to reproduce the strong differential absorption signal observed at 62.5\,eV in the experiment, a dark state (d\textsubscript{2}) located around 63.3 eV (one NIR photon above X\textsubscript{2$''$}) is introduced to the model system, and this state is reached by the NIR pulse. To achieve better agreement with the experimental results, d\textsubscript{2} was also coupled to X\textsubscript{2$'$} and \3. The observed differential absorption signal at 59.2\,eV indicates that it is also necessary to consider a lower-energy dark state d\textsubscript{1} around 60.80\,eV to fully capture the NIR-induced core exciton dynamics observed in the experimental results. The values of the precise energies of the two dark states, the transition dipole moments describing the NIR-induced couplings between the bright and dark states, and the AC Stark phase constant are allowed to freely vary and are optimized to best match the experimental results obtained in the temporally-stabilized measurement through least-squares minimization. The values obtained through least-squares minimization of the few level model with respect to the experimental temporally-stabilized transient absorption data (Fig. \ref{Stabi}(a) of the main text) are outlined in Table \ref{FullSim}.\par
\begin{figure}
\includegraphics{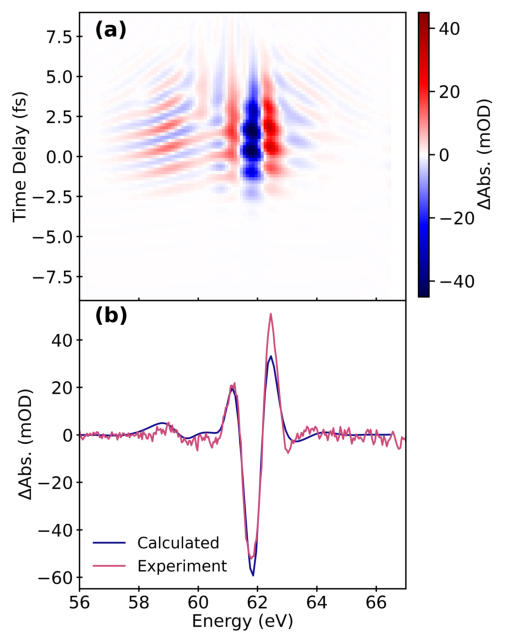}%
\caption{\label{Simulation}(a) Simulated transient absorption spectrum obtained using parameters in Table \ref{FullSim}. (b) Experimental stabilized (red curve) and simulated (blue curve) differential absorption spectra taken at $\tau$\,=\,1\,fs.}
\end{figure}

\begin{table} 
\caption{\label{FullSim} Optimized parameters and confidence intervals (C.I.) for the parameters used in simulating the LiF core exciton dynamics. All transition dipole moments $\mu$\textit{\textsubscript{p,q}} are given in atomic units. $\alpha$ is unitless.}
\begin{ruledtabular}
\begin{tabular}{ c c c }
Parameter & Value & C.I. \\
\hline
d\textsubscript{1} (eV) & 60.83 & 0.03\\
$\mu${\textsubscript{1,d\textsubscript{1}}} & 5.67 & 1.51\\
$\mu${\textsubscript{2$'$,d\textsubscript{1}}} & -1.90 & 0.30\\
$\mu${\textsubscript{2$''$,d\textsubscript{1}}} & 0.59 & 0.16\\
d\textsubscript{2} (eV) & 63.23 & 0.04\\
$\mu${\textsubscript{2$'$,d\textsubscript{2}}} & 0.29 & 0.09\\
$\mu${\textsubscript{2$''$,d\textsubscript{2}}} & 2.20 & 0.05\\
$\mu${\textsubscript{3,d\textsubscript{2}}} & 0.025 & 0.13\\
$\alpha$ & 0.40 & 0.15\\
\end{tabular}
\end{ruledtabular}
\end{table}
The simulated transient absorption spectrum using these parameters is presented in Fig. \ref{Simulation}(a), and comparison of the simulation to the temporally-stabilized experimental result at $\tau$\,=\,1\,fs is given in Fig, \ref{Simulation}(b).
The simulated data is normalized to the scale of the experimental result obtained when using parallel beam polarizations and temporal stabilization. The chi-square minimization of the simulation gives a value of 0.00232, demonstrating a close agreement between the calculated and measured results. The calculation is able to reproduce the characteristic profile observed in the experiment between 61 and 63 eV, as well as the broad, low-amplitude signal at 59.25 eV.\par
\begin{figure}
\includegraphics{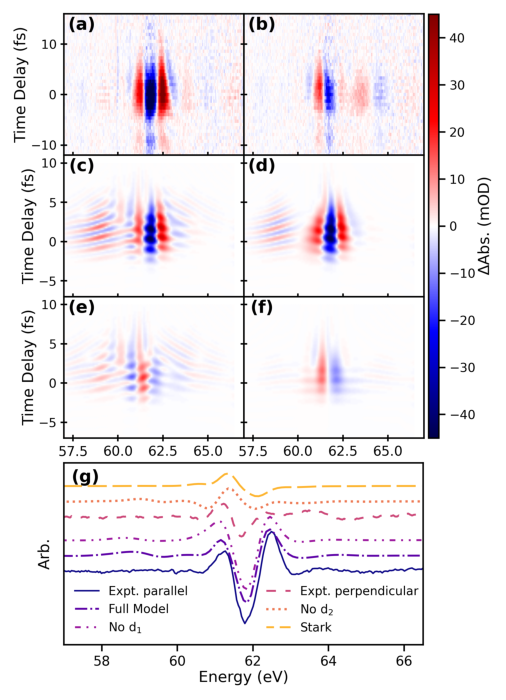}%
\caption{\label{PartialSim} Panels (a) and (b) show the experimental transient absorption result obtained when using parallel (a) and perpendicular (b) XUV and NIR beam polarizations. Panels (c)-(f) show the simulated transient absorption spectra obtained from the model by varying the inclusion of dark states d\textsubscript{1} and d\textsubscript{2}. (c) presents the full model, incorporating all electronic states as detailed in Table \ref{FullSim}. (d) The simulation when excluding d\textsubscript{1}. (e) The simulation when excluding d\textsubscript{2}. (f) The simulation when excluding both d\textsubscript{1} and d\textsubscript{2}. Panel (g) gives the transient absorption signal for (a)-(f) taken at $\tau$\,=\,1\,fs.}
\end{figure}
The individual roles of the dark states are illustrated in Fig. \ref{PartialSim}. Panels (a) and (b) show the experimental result obtained when using parallel and perpendicular XUV and NIR polarizations, respectively, included here for comparison to the variations of the model system. (c) shows the simulated delay-dependent absorption spectrum for the full model, using the parameters outlined in Table \ref{FullSim}. As shown in Fig. \ref{Simulation}, the simulated data agrees well with the experimental results. To aid in comparison, panel (g) shows the differential absorption of (a)-(f) taken at $\tau$\,=\,1\,fs. In panels (d) and (e) we isolate the roles of the different dark states by turning off the coupling to d\textsubscript{1} and d\textsubscript{2}, respectively. In panel (d), the characteristic profile of strong positive (60.5, 62.5\,eV) and negative (61.7\,eV) peaks observed experimentally between 61 and 63\,eV persists. This feature arises from the strong NIR-resonant coupling between \2\textsubscript{$''$} and d\textsubscript{2}, and is thus relatively unaffected by the absence of couplings to d\textsubscript{1}. The positive feature at 61\,eV is found to broaden in the absence of d\textsubscript{1}, extending to lower photon energies than in the experimental measurement. The broad feature centered just below 59\,eV (0.2\,eV lower than in the experiment) is due to two-NIR-photon couplings to \2\textsubscript{$''$}. In contrast, when the coupling to d\textsubscript{2} is turned off and couplings to d\textsubscript{1} are preserved (panel (e)), the characteristic peak pattern from 61-63\,eV observed in (a) disappears and is replaced by a weaker feature with a more complex structure. The low-energy feature centered on 59.2\,eV is a light-induced state facilitated by one-NIR-photon couplings to d\textsubscript{1}. These observations suggest that the broad feature around 59.2\,eV in the experimental data arises from the combined influence of dark states d\textsubscript{1} and d\textsubscript{2}. Finally, Fig. \ref{PartialSim}(f) displays the simulated transient absorption spectrum when both dark states are excluded from the model. In this case, the AC Stark phase is the only delay-dependent dynamic phase remaining in the calculations. Here, both the feature around 59\,eV and the characteristic differential absorption profile are absent. The resulting shape is in qualitative agreement with the experimental line shape obtained using crossed beam polarizations, suggesting that the coupling to the dark states is severely dampened when the XUV and NIR polarizations are not parallel. It is important to be noted again that panel (f) is not obtained by fitting the experimental result when using perpendicular beam polarizations (panel (b)), but is instead simply the result of excluding the dark states d\textsubscript{1} and d\textsubscript{2} from the model obtained by optimization with the experiment parallel beam polarization measurement. It is then unsurprising that there is a deviation in absolute peak heights and central frequencies between the experimental crossed polarization result and the model including only the Stark shifts of the bright states. For the result illustrated in panel (f) to exactly replicate experimental crossed polarization results, there would have to be a complete elimination of all couplings between the bright and dark states in the experiment. As noted in the main text, the transient absorption signals associated with the coupling between \2 and \3 is suppressed by approximately 90\%, indicating there is residual coupling in the experiment not captured by the pure Stark shifts described in Fig. \ref{PartialSim}(f). However, the significant similarities between these two results indicate that the transient absorption signals observed when using crossed XUV and NIR polarizations in the experiment are most likely dominated by Stark shifts of the bright states, in particular the 1\s2\p\ bright exciton. 
\subsection{\label{DFT Methods} Density Functional Theory Calculations}
\begin{figure}
\includegraphics{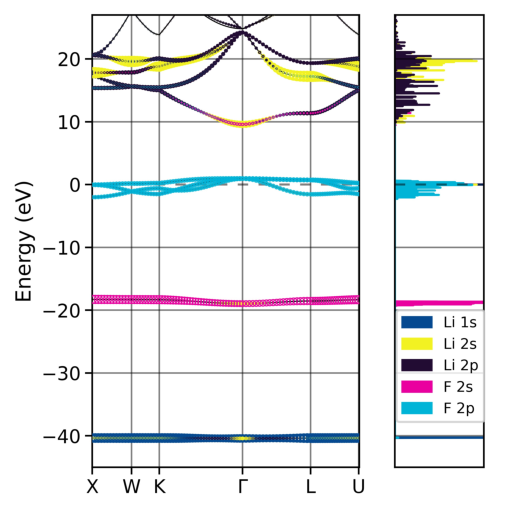}%
\caption{\label{FullBand} LiF band structure and projected density of states calculated using Quantum Espresso, showing the orbital character of the core, valence, and conduction bands.}
\end{figure}

The methods used to calculate the ground and excited state projected band structures in LiF closely follow those described by Quintero-Bermudez and Leone \cite{quintero-bermudez_deciphering_2024}. An 8\,x\,8\,x\,8 \textit{k} point mesh and Perdew-Burke-Ernzerhof (PBE) projector augmented wave (KJPAW) functionals are used for calculations of the LiF projected band structure in Quantum Espresso \cite{giannozzi_advanced_2017, giannozzi_quantum_2009}. The size of the \textit{k} point mesh was optimized through convergence tests. The full calculated band structure and pDOS, including the Li 1\s\ and F 2\s\ core levels, is given in Fig. \ref{FullBand}. The calculated bandgap ($\sim$7\,eV) is nearly half of the experimental bandgap energy of 13.6\,eV, as is common for DFT calculations \cite{MoriSanchez2008,Perdew1983,Sham1983}. We do not attempt to correct the bandgap energies of the band structures presented here, as the aim of these calculations is to interrogate the orbital character of the excited core exciton states, rather than their precise energies. \par
Calculations of the exciton weight projections are performed using the Exciting software package \cite{gulans_exciting_2014,draxl_exciting_2017,vorwerk_addressing_2017}. Ground state calculations using PBE generalized gradient approximation functionals and an 8\,x\,8\,x\,8 \textit{k} point mesh are first performed. The Bethe-Salpeter equation (BSE) is then used to calculate excited state properties, including the calculated XUV absorption spectrum shown in the main text. The BSE calculation uses 5 occupied atomic levels (Li 1\s, F 2\s, three F 2\p) and 5 unoccupied conduction band levels to calculate the excited states lying between 60 and 72\,eV. The BSE-calculated excitonic wavefunctions are then visualized in \textit{k} space by projection onto a sum of the ground state valence and conduction band levels. In Fig. \ref{Weights}(a) of the main text, the 14\,eV scissor correction is applied while plotting to obtain agreement in the energy of the calculated and experimental \2 peaks. The 14\,eV arises from the $\sim$7\,eV underestimation of the bandgap, and a further 7\,eV underestimation of the Li 1s to CB transition energy. The correction is not applied during the BSE calculations, and as such is not reflected in the band structures depicting the excitonic weights.\par

\section{Experimental Data}
\begin{figure}
\includegraphics{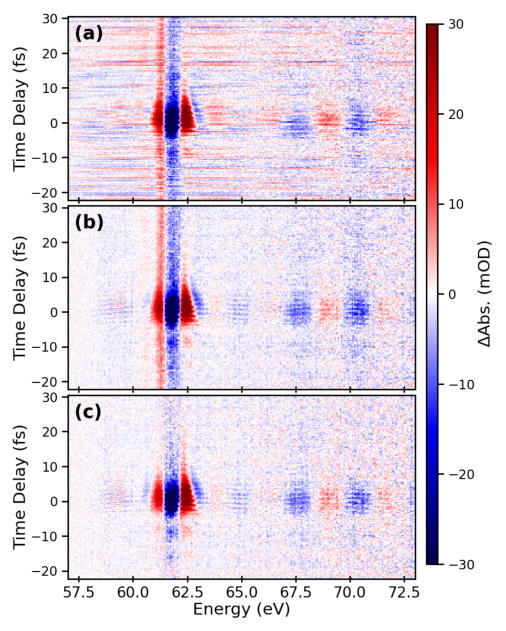}%
\caption{\label{Stabilized} Experimental transient absorption data for delay-stabilized measurements of LiF, for the full range of -22 to +30\,fs. (a) Untreated experimental data. (b) Edge referenced data. (c) Edge referenced data with heat background subtracted, as presented in the main text.}
\end{figure}

\begin{figure}
\includegraphics{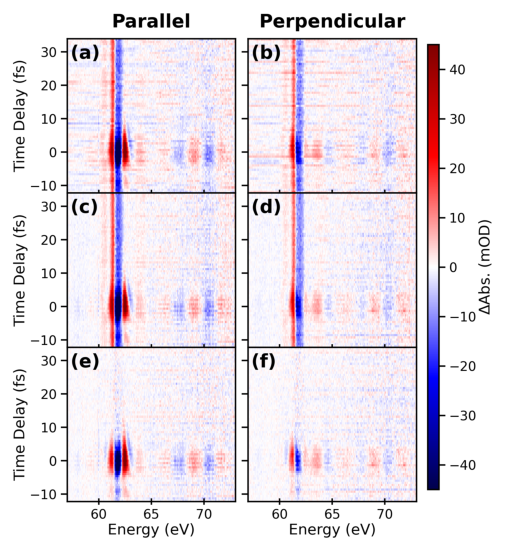}%
\caption{\label{Polarization} Transient absorption data for parallel (a),(c),(e) and perpendicular (b),(d),(f) XUV and NIR polarizations, for the delay range of -12 to +33\,fs. The untreated data is given in (a),(b), the spectrum after edge referencing is given in (c),(d), and the heat-subtracted edge referenced data as presented in the main text is in (e),(f).}
\end{figure}

The untreated differential absorption of LiF obtained by temporally stabilizing the delay of the 6.8\,x\,10\textsuperscript{12}\,W/cm\textsuperscript{2} NIR probe pulse is presented in Fig. \ref{Stabilized}(a).
Edge referencing \cite{geneaux_source_2021} is then implemented to reduce the considerable noise background due to fluctuations in the XUV pump intensity (Fig. \ref{Stabilized}(b)). The range of 35-55\,eV was chosen as the reference, as no XUV absorption features are present here. Finally, the residual heat background, present at large negative time delays and caused by laser heating of the sample, is subtracted (Fig. \ref{Stabilized}(c)). The heat-subtracted form of the data is then used for further analysis. The same signal processing procedure was applied to the polarization-dependent data, presented here in Fig. \ref{Polarization}.
\bibliography{refs}

\begin{thebibliography}{76}%
\makeatletter
\providecommand \@ifxundefined [1]{%
 \@ifx{#1\undefined}
}%
\providecommand \@ifnum [1]{%
 \ifnum #1\expandafter \@firstoftwo
 \else \expandafter \@secondoftwo
 \fi
}%
\providecommand \@ifx [1]{%
 \ifx #1\expandafter \@firstoftwo
 \else \expandafter \@secondoftwo
 \fi
}%
\providecommand \natexlab [1]{#1}%
\providecommand \enquote  [1]{``#1''}%
\providecommand \bibnamefont  [1]{#1}%
\providecommand \bibfnamefont [1]{#1}%
\providecommand \citenamefont [1]{#1}%
\providecommand \href@noop [0]{\@secondoftwo}%
\providecommand \href [0]{\begingroup \@sanitize@url \@href}%
\providecommand \@href[1]{\@@startlink{#1}\@@href}%
\providecommand \@@href[1]{\endgroup#1\@@endlink}%
\providecommand \@sanitize@url [0]{\catcode `\\12\catcode `\$12\catcode `\&12\catcode `\#12\catcode `\^12\catcode `\_12\catcode `\%12\relax}%
\providecommand \@@startlink[1]{}%
\providecommand \@@endlink[0]{}%
\providecommand \url  [0]{\begingroup\@sanitize@url \@url }%
\providecommand \@url [1]{\endgroup\@href {#1}{\urlprefix }}%
\providecommand \urlprefix  [0]{URL }%
\providecommand \Eprint [0]{\href }%
\providecommand \doibase [0]{https://doi.org/}%
\providecommand \selectlanguage [0]{\@gobble}%
\providecommand \bibinfo  [0]{\@secondoftwo}%
\providecommand \bibfield  [0]{\@secondoftwo}%
\providecommand \translation [1]{[#1]}%
\providecommand \BibitemOpen [0]{}%
\providecommand \bibitemStop [0]{}%
\providecommand \bibitemNoStop [0]{.\EOS\space}%
\providecommand \EOS [0]{\spacefactor3000\relax}%
\providecommand \BibitemShut  [1]{\csname bibitem#1\endcsname}%
\let\auto@bib@innerbib\@empty
\bibitem [{\citenamefont {Bussert}\ \emph {et~al.}(1987)\citenamefont {Bussert}, \citenamefont {Neuschäfer},\ and\ \citenamefont {Leone}}]{bussert_effect_1987}%
  \BibitemOpen
  \bibfield  {author} {\bibinfo {author} {\bibfnamefont {W.}~\bibnamefont {Bussert}}, \bibinfo {author} {\bibfnamefont {D.}~\bibnamefont {Neuschäfer}},\ and\ \bibinfo {author} {\bibfnamefont {S.}~\bibnamefont {Leone}},\ }\bibfield  {title} {\bibinfo {title} {The effect of orbital alignment on the forward and reverse electronic energy transfer {Ca}(4s5p \textsuperscript{1}{P}\textsubscript{1})+{M}$\rightleftarrows${Ca}(4s5p \textsuperscript{3}{P}\textsubscript{J})+{M} with rare gases},\ }\href {https://doi.org/10.1063/1.452938} {\bibfield  {journal} {\bibinfo  {journal} {J. Chem. Phys.}\ }\textbf {\bibinfo {volume} {87}},\ \bibinfo {pages} {3833} (\bibinfo {year} {1987})}\BibitemShut {NoStop}%
\bibitem [{\citenamefont {Driessen}\ \emph {et~al.}(1991{\natexlab{a}})\citenamefont {Driessen}, \citenamefont {Smith},\ and\ \citenamefont {Leone}}]{driessen_alignment_1991}%
  \BibitemOpen
  \bibfield  {author} {\bibinfo {author} {\bibfnamefont {J.~P.~J.}\ \bibnamefont {Driessen}}, \bibinfo {author} {\bibfnamefont {C.~J.}\ \bibnamefont {Smith}},\ and\ \bibinfo {author} {\bibfnamefont {S.~R.}\ \bibnamefont {Leone}},\ }\bibfield  {title} {\bibinfo {title} {{Alignment effects of \ensuremath{\Vert}J=3〉 states prepared by three-photon excitation: Sixfold Symmetry in Collisional Energy Transfer, Ca(4s4f${,}^{1}$${\mathit{F}}_{3}$)+He\ensuremath{\rightarrow}Ca(4${\mathit{p}}^{2}$${,}^{1}$${\mathit{S}}_{0}$)+He}},\ }\href {https://doi.org/10.1103/PhysRevA.44.R1431} {\bibfield  {journal} {\bibinfo  {journal} {Phys. Rev. A}\ }\textbf {\bibinfo {volume} {44}},\ \bibinfo {pages} {R1431} (\bibinfo {year} {1991}{\natexlab{a}})}\BibitemShut {NoStop}%
\bibitem [{\citenamefont {Driessen}\ \emph {et~al.}(1991{\natexlab{b}})\citenamefont {Driessen}, \citenamefont {Smith},\ and\ \citenamefont {Leone}}]{driessen_relative_1991}%
  \BibitemOpen
  \bibfield  {author} {\bibinfo {author} {\bibfnamefont {J.~P.~J.}\ \bibnamefont {Driessen}}, \bibinfo {author} {\bibfnamefont {C.~J.}\ \bibnamefont {Smith}},\ and\ \bibinfo {author} {\bibfnamefont {S.~R.}\ \bibnamefont {Leone}},\ }\bibfield  {title} {\bibinfo {title} {{Relative cross sections for calcium({\textsuperscript{1}F\textsubscript{3}}) sublevels ({\textbar}{M}{\textbar} = 0-3) through alignment effects in collisional energy transfer: Calcium(4s4f,{\textsuperscript{1}F\textsubscript{3}}) + rare gas \textrightarrow\ Calcium(4p\textsuperscript{2},{\textsuperscript{1}S\textsubscript{0}}) + rare gas as a function of rare gas}},\ }\href {https://doi.org/10.1021/j100174a027} {\bibfield  {journal} {\bibinfo  {journal} {J. Phys. Chem.}\ }\textbf {\bibinfo {volume} {95}},\ \bibinfo {pages} {8163} (\bibinfo {year} {1991}{\natexlab{b}})}\BibitemShut {NoStop}%
\bibitem [{\citenamefont {Driessen}\ and\ \citenamefont {Leone}(1992)}]{driessen_n-vector_1992}%
  \BibitemOpen
  \bibfield  {author} {\bibinfo {author} {\bibfnamefont {J.~P.~J.}\ \bibnamefont {Driessen}}\ and\ \bibinfo {author} {\bibfnamefont {S.~R.}\ \bibnamefont {Leone}},\ }\bibfield  {title} {\bibinfo {title} {{n-Vector correlations in collision dynamics with atomic orbital alignment: the importance of coherence denoting azimuthal structure for n $\ge$ 3}},\ }\href {https://doi.org/10.1021/j100194a012} {\bibfield  {journal} {\bibinfo  {journal} {J. Phys. Chem.}\ }\textbf {\bibinfo {volume} {96}},\ \bibinfo {pages} {6136} (\bibinfo {year} {1992})}\BibitemShut {NoStop}%
\bibitem [{\citenamefont {Smith}\ \emph {et~al.}(1992)\citenamefont {Smith}, \citenamefont {Driessen}, \citenamefont {Eno},\ and\ \citenamefont {Leone}}]{smith_laser_1992}%
  \BibitemOpen
  \bibfield  {author} {\bibinfo {author} {\bibfnamefont {C.~J.}\ \bibnamefont {Smith}}, \bibinfo {author} {\bibfnamefont {J.~P.~J.}\ \bibnamefont {Driessen}}, \bibinfo {author} {\bibfnamefont {L.}~\bibnamefont {Eno}},\ and\ \bibinfo {author} {\bibfnamefont {S.~R.}\ \bibnamefont {Leone}},\ }\bibfield  {title} {\bibinfo {title} {Laser preparation and probing of initial and final orbital alignment in collision‐induced energy transfer {Ca}(4s5p,\textsuperscript{1}{P}\textsubscript{1}) +{He}→{Ca}(4s5p,\textsuperscript{3}{P}\textsubscript{2})+{He}},\ }\href {https://doi.org/10.1063/1.462326} {\bibfield  {journal} {\bibinfo  {journal} {J. Chem. Phys.}\ }\textbf {\bibinfo {volume} {96}},\ \bibinfo {pages} {8212} (\bibinfo {year} {1992})}\BibitemShut {NoStop}%
\bibitem [{\citenamefont {de~Vivie‐Riedle}\ \emph {et~al.}(1993)\citenamefont {de~Vivie‐Riedle}, \citenamefont {Driessen},\ and\ \citenamefont {Leone}}]{de_vivieriedle_threevector_1993}%
  \BibitemOpen
  \bibfield  {author} {\bibinfo {author} {\bibfnamefont {R.}~\bibnamefont {de~Vivie‐Riedle}}, \bibinfo {author} {\bibfnamefont {J.~P.~J.}\ \bibnamefont {Driessen}},\ and\ \bibinfo {author} {\bibfnamefont {S.~R.}\ \bibnamefont {Leone}},\ }\bibfield  {title} {\bibinfo {title} {{Three‐vector correlation study of orientation and coherence effects in {Na}(3p,{\textsuperscript{2}P}\textsubscript{1/2}←{\textsuperscript{2}P}\textsubscript{3/2})+{He}: Semiclassical and quantum calculations}},\ }\href {https://doi.org/10.1063/1.465053} {\bibfield  {journal} {\bibinfo  {journal} {J. Chem. Phys.}\ }\textbf {\bibinfo {volume} {98}},\ \bibinfo {pages} {2038} (\bibinfo {year} {1993})}\BibitemShut {NoStop}%
\bibitem [{\citenamefont {Smith}\ \emph {et~al.}(1993)\citenamefont {Smith}, \citenamefont {Spain}, \citenamefont {Dalberth}, \citenamefont {Leone},\ and\ \citenamefont {Driessen}}]{smith_initial_1993}%
  \BibitemOpen
  \bibfield  {author} {\bibinfo {author} {\bibfnamefont {C.~J.}\ \bibnamefont {Smith}}, \bibinfo {author} {\bibfnamefont {E.~M.}\ \bibnamefont {Spain}}, \bibinfo {author} {\bibfnamefont {M.~J.}\ \bibnamefont {Dalberth}}, \bibinfo {author} {\bibfnamefont {S.~R.}\ \bibnamefont {Leone}},\ and\ \bibinfo {author} {\bibfnamefont {J.~P.~J.}\ \bibnamefont {Driessen}},\ }\bibfield  {title} {\bibinfo {title} {Initial and final orbital alignment probing of the fine-structure-changing collisions among the {Ca} (4s)\textsuperscript{1}(4p)\textsuperscript{1}, \textsuperscript{3}{P}\textsubscript{J} states with {He}: determination of coherence and conventional cross-sections},\ }\href {https://doi.org/10.1039/FT9938901401} {\bibfield  {journal} {\bibinfo  {journal} {J. Chem. Soc., Faraday Trans.}\ }\textbf {\bibinfo {volume} {89}},\ \bibinfo {pages} {1401} (\bibinfo {year} {1993})}\BibitemShut {NoStop}%
\bibitem [{\citenamefont {Spain}\ \emph {et~al.}(1995)\citenamefont {Spain}, \citenamefont {Dalberth}, \citenamefont {Kleiber}, \citenamefont {Leone}, \citenamefont {Op~De~Beek},\ and\ \citenamefont {Driessen}}]{spain_orbital_1995}%
  \BibitemOpen
  \bibfield  {author} {\bibinfo {author} {\bibfnamefont {E.~M.}\ \bibnamefont {Spain}}, \bibinfo {author} {\bibfnamefont {M.~J.}\ \bibnamefont {Dalberth}}, \bibinfo {author} {\bibfnamefont {P.~D.}\ \bibnamefont {Kleiber}}, \bibinfo {author} {\bibfnamefont {S.~R.}\ \bibnamefont {Leone}}, \bibinfo {author} {\bibfnamefont {S.~S.}\ \bibnamefont {Op~De~Beek}},\ and\ \bibinfo {author} {\bibfnamefont {J.~P.~J.}\ \bibnamefont {Driessen}},\ }\bibfield  {title} {\bibinfo {title} {Orbital alignment cross sections by stimulated emission probing: {The} state-to-state {Ca} {Rydberg} process {Ca}(4\textit{s}17\textit{d} \textsuperscript{1}\textit{{D}}\textsubscript{2})+{Xe}→{Ca}(4\textit{s}18\textit{p} \textsuperscript{1}\textit{{P}}\textsubscript{1})+{Xe}},\ }\href {https://doi.org/10.1063/1.468768} {\bibfield  {journal} {\bibinfo  {journal} {J. Chem. Phys.}\ }\textbf {\bibinfo {volume} {102}},\ \bibinfo {pages} {9532} (\bibinfo {year} {1995})}\BibitemShut {NoStop}%
\bibitem [{\citenamefont {Zeidler}\ \emph {et~al.}(2005)\citenamefont {Zeidler}, \citenamefont {Staudte}, \citenamefont {Bardon}, \citenamefont {Villeneuve}, \citenamefont {D\"orner},\ and\ \citenamefont {Corkum}}]{zeidler_controlling_2005}%
  \BibitemOpen
  \bibfield  {author} {\bibinfo {author} {\bibfnamefont {D.}~\bibnamefont {Zeidler}}, \bibinfo {author} {\bibfnamefont {A.}~\bibnamefont {Staudte}}, \bibinfo {author} {\bibfnamefont {A.~B.}\ \bibnamefont {Bardon}}, \bibinfo {author} {\bibfnamefont {D.~M.}\ \bibnamefont {Villeneuve}}, \bibinfo {author} {\bibfnamefont {R.}~\bibnamefont {D\"orner}},\ and\ \bibinfo {author} {\bibfnamefont {P.~B.}\ \bibnamefont {Corkum}},\ }\bibfield  {title} {\bibinfo {title} {{Controlling Attosecond Double Ionization Dynamics via Molecular Alignment}},\ }\href {https://doi.org/10.1103/PhysRevLett.95.203003} {\bibfield  {journal} {\bibinfo  {journal} {Phys. Rev. Lett.}\ }\textbf {\bibinfo {volume} {95}},\ \bibinfo {pages} {203003} (\bibinfo {year} {2005})}\BibitemShut {NoStop}%
\bibitem [{\citenamefont {Reduzzi}\ \emph {et~al.}(2015)\citenamefont {Reduzzi}, \citenamefont {Hummert}, \citenamefont {Dubrouil}, \citenamefont {Calegari}, \citenamefont {Nisoli}, \citenamefont {Frassetto}, \citenamefont {Poletto}, \citenamefont {Chen}, \citenamefont {Wu}, \citenamefont {Gaarde}, \citenamefont {Schafer},\ and\ \citenamefont {Sansone}}]{reduzzi_polarization_2015}%
  \BibitemOpen
  \bibfield  {author} {\bibinfo {author} {\bibfnamefont {M.}~\bibnamefont {Reduzzi}}, \bibinfo {author} {\bibfnamefont {J.}~\bibnamefont {Hummert}}, \bibinfo {author} {\bibfnamefont {A.}~\bibnamefont {Dubrouil}}, \bibinfo {author} {\bibfnamefont {F.}~\bibnamefont {Calegari}}, \bibinfo {author} {\bibfnamefont {M.}~\bibnamefont {Nisoli}}, \bibinfo {author} {\bibfnamefont {F.}~\bibnamefont {Frassetto}}, \bibinfo {author} {\bibfnamefont {L.}~\bibnamefont {Poletto}}, \bibinfo {author} {\bibfnamefont {S.}~\bibnamefont {Chen}}, \bibinfo {author} {\bibfnamefont {M.}~\bibnamefont {Wu}}, \bibinfo {author} {\bibfnamefont {M.~B.}\ \bibnamefont {Gaarde}}, \bibinfo {author} {\bibfnamefont {K.}~\bibnamefont {Schafer}},\ and\ \bibinfo {author} {\bibfnamefont {G.}~\bibnamefont {Sansone}},\ }\bibfield  {title} {\bibinfo {title} {Polarization control of absorption of virtual dressed states in helium},\ }\href {https://doi.org/10.1103/PhysRevA.92.033408} {\bibfield  {journal} {\bibinfo  {journal} {Phys. Rev. A}\ }\textbf
  {\bibinfo {volume} {92}},\ \bibinfo {pages} {033408} (\bibinfo {year} {2015})}\BibitemShut {NoStop}%
\bibitem [{\citenamefont {Chew}\ \emph {et~al.}(2018)\citenamefont {Chew}, \citenamefont {Douguet}, \citenamefont {Cariker}, \citenamefont {Li}, \citenamefont {Lindroth}, \citenamefont {Ren}, \citenamefont {Yin}, \citenamefont {Argenti}, \citenamefont {Hill},\ and\ \citenamefont {Chang}}]{chew_attosecond_2018}%
  \BibitemOpen
  \bibfield  {author} {\bibinfo {author} {\bibfnamefont {A.}~\bibnamefont {Chew}}, \bibinfo {author} {\bibfnamefont {N.}~\bibnamefont {Douguet}}, \bibinfo {author} {\bibfnamefont {C.}~\bibnamefont {Cariker}}, \bibinfo {author} {\bibfnamefont {J.}~\bibnamefont {Li}}, \bibinfo {author} {\bibfnamefont {E.}~\bibnamefont {Lindroth}}, \bibinfo {author} {\bibfnamefont {X.}~\bibnamefont {Ren}}, \bibinfo {author} {\bibfnamefont {Y.}~\bibnamefont {Yin}}, \bibinfo {author} {\bibfnamefont {L.}~\bibnamefont {Argenti}}, \bibinfo {author} {\bibfnamefont {W.~T.}\ \bibnamefont {Hill}},\ and\ \bibinfo {author} {\bibfnamefont {Z.}~\bibnamefont {Chang}},\ }\bibfield  {title} {\bibinfo {title} {Attosecond transient absorption spectrum of argon at the ${L}_{2,3}$ edge},\ }\href {https://doi.org/10.1103/PhysRevA.97.031407} {\bibfield  {journal} {\bibinfo  {journal} {Phys. Rev. A}\ }\textbf {\bibinfo {volume} {97}},\ \bibinfo {pages} {031407} (\bibinfo {year} {2018})}\BibitemShut {NoStop}%
\bibitem [{\citenamefont {Pandey}\ \emph {et~al.}(2023)\citenamefont {Pandey}, \citenamefont {Min}, \citenamefont {Reddeppa}, \citenamefont {Malhotra}, \citenamefont {Xiao}, \citenamefont {Wu}, \citenamefont {Sun},\ and\ \citenamefont {Mi}}]{pandey_ultrahigh_2023}%
  \BibitemOpen
  \bibfield  {author} {\bibinfo {author} {\bibfnamefont {A.}~\bibnamefont {Pandey}}, \bibinfo {author} {\bibfnamefont {J.}~\bibnamefont {Min}}, \bibinfo {author} {\bibfnamefont {M.}~\bibnamefont {Reddeppa}}, \bibinfo {author} {\bibfnamefont {Y.}~\bibnamefont {Malhotra}}, \bibinfo {author} {\bibfnamefont {Y.}~\bibnamefont {Xiao}}, \bibinfo {author} {\bibfnamefont {Y.}~\bibnamefont {Wu}}, \bibinfo {author} {\bibfnamefont {K.}~\bibnamefont {Sun}},\ and\ \bibinfo {author} {\bibfnamefont {Z.}~\bibnamefont {Mi}},\ }\bibfield  {title} {\bibinfo {title} {An {Ultrahigh} {Efficiency} {Excitonic} {Micro}-{LED}},\ }\href {https://doi.org/10.1021/acs.nanolett.2c04220} {\bibfield  {journal} {\bibinfo  {journal} {Nano Lett.}\ }\textbf {\bibinfo {volume} {23}},\ \bibinfo {pages} {1680} (\bibinfo {year} {2023})}\BibitemShut {NoStop}%
\bibitem [{\citenamefont {Gregg}(2003)}]{gregg_excitonic_2003}%
  \BibitemOpen
  \bibfield  {author} {\bibinfo {author} {\bibfnamefont {B.~A.}\ \bibnamefont {Gregg}},\ }\bibfield  {title} {\bibinfo {title} {Excitonic {Solar} {Cells}},\ }\href {https://doi.org/10.1021/jp022507x} {\bibfield  {journal} {\bibinfo  {journal} {J. Phys. Chem. B}\ }\textbf {\bibinfo {volume} {107}},\ \bibinfo {pages} {4688} (\bibinfo {year} {2003})}\BibitemShut {NoStop}%
\bibitem [{\citenamefont {Menke}\ and\ \citenamefont {Holmes}(2014)}]{menke_exciton_2014}%
  \BibitemOpen
  \bibfield  {author} {\bibinfo {author} {\bibfnamefont {S.~M.}\ \bibnamefont {Menke}}\ and\ \bibinfo {author} {\bibfnamefont {R.~J.}\ \bibnamefont {Holmes}},\ }\bibfield  {title} {\bibinfo {title} {Exciton diffusion in organic photovoltaic cells},\ }\href {https://doi.org/10.1039/C3EE42444H} {\bibfield  {journal} {\bibinfo  {journal} {Energy Environ. Sci.}\ }\textbf {\bibinfo {volume} {7}},\ \bibinfo {pages} {499} (\bibinfo {year} {2014})}\BibitemShut {NoStop}%
\bibitem [{\citenamefont {Classen}\ \emph {et~al.}(2020)\citenamefont {Classen}, \citenamefont {Chochos}, \citenamefont {Lüer}, \citenamefont {Gregoriou}, \citenamefont {Wortmann}, \citenamefont {Osvet}, \citenamefont {Forberich}, \citenamefont {McCulloch}, \citenamefont {Heumüller},\ and\ \citenamefont {Brabec}}]{classen_role_2020}%
  \BibitemOpen
  \bibfield  {author} {\bibinfo {author} {\bibfnamefont {A.}~\bibnamefont {Classen}}, \bibinfo {author} {\bibfnamefont {C.~L.}\ \bibnamefont {Chochos}}, \bibinfo {author} {\bibfnamefont {L.}~\bibnamefont {Lüer}}, \bibinfo {author} {\bibfnamefont {V.~G.}\ \bibnamefont {Gregoriou}}, \bibinfo {author} {\bibfnamefont {J.}~\bibnamefont {Wortmann}}, \bibinfo {author} {\bibfnamefont {A.}~\bibnamefont {Osvet}}, \bibinfo {author} {\bibfnamefont {K.}~\bibnamefont {Forberich}}, \bibinfo {author} {\bibfnamefont {I.}~\bibnamefont {McCulloch}}, \bibinfo {author} {\bibfnamefont {T.}~\bibnamefont {Heumüller}},\ and\ \bibinfo {author} {\bibfnamefont {C.~J.}\ \bibnamefont {Brabec}},\ }\bibfield  {title} {\bibinfo {title} {The role of exciton lifetime for charge generation in organic solar cells at negligible energy-level offsets},\ }\href {https://doi.org/10.1038/s41560-020-00684-7} {\bibfield  {journal} {\bibinfo  {journal} {Nat. Energy}\ }\textbf {\bibinfo {volume} {5}},\ \bibinfo {pages} {711} (\bibinfo {year}
  {2020})}\BibitemShut {NoStop}%
\bibitem [{\citenamefont {Zhu}\ \emph {et~al.}(2022)\citenamefont {Zhu}, \citenamefont {Wei},\ and\ \citenamefont {Yi}}]{zhu_exciton_2022}%
  \BibitemOpen
  \bibfield  {author} {\bibinfo {author} {\bibfnamefont {L.}~\bibnamefont {Zhu}}, \bibinfo {author} {\bibfnamefont {Z.}~\bibnamefont {Wei}},\ and\ \bibinfo {author} {\bibfnamefont {Y.}~\bibnamefont {Yi}},\ }\bibfield  {title} {\bibinfo {title} {Exciton {Binding} {Energies} in {Organic} {Photovoltaic} {Materials}: {A} {Theoretical} {Perspective}},\ }\href {https://doi.org/10.1021/acs.jpcc.1c08898} {\bibfield  {journal} {\bibinfo  {journal} {J. Phys. Chem. C}\ }\textbf {\bibinfo {volume} {126}},\ \bibinfo {pages} {14} (\bibinfo {year} {2022})}\BibitemShut {NoStop}%
\bibitem [{\citenamefont {Ghosh}\ and\ \citenamefont {Liew}(2020)}]{ghosh_quantum_2020}%
  \BibitemOpen
  \bibfield  {author} {\bibinfo {author} {\bibfnamefont {S.}~\bibnamefont {Ghosh}}\ and\ \bibinfo {author} {\bibfnamefont {T.~C.~H.}\ \bibnamefont {Liew}},\ }\bibfield  {title} {\bibinfo {title} {Quantum computing with exciton-polariton condensates},\ }\href {https://doi.org/10.1038/s41534-020-0244-x} {\bibfield  {journal} {\bibinfo  {journal} {Npj Quantum Inf.}\ }\textbf {\bibinfo {volume} {6}},\ \bibinfo {pages} {1} (\bibinfo {year} {2020})}\BibitemShut {NoStop}%
\bibitem [{\citenamefont {Harankahage}\ \emph {et~al.}(2021)\citenamefont {Harankahage}, \citenamefont {Cassidy}, \citenamefont {Yang}, \citenamefont {Porotnikov}, \citenamefont {Williams}, \citenamefont {Kholmicheva},\ and\ \citenamefont {Zamkov}}]{harankahage_quantum_2021}%
  \BibitemOpen
  \bibfield  {author} {\bibinfo {author} {\bibfnamefont {D.}~\bibnamefont {Harankahage}}, \bibinfo {author} {\bibfnamefont {J.}~\bibnamefont {Cassidy}}, \bibinfo {author} {\bibfnamefont {M.}~\bibnamefont {Yang}}, \bibinfo {author} {\bibfnamefont {D.}~\bibnamefont {Porotnikov}}, \bibinfo {author} {\bibfnamefont {M.}~\bibnamefont {Williams}}, \bibinfo {author} {\bibfnamefont {N.}~\bibnamefont {Kholmicheva}},\ and\ \bibinfo {author} {\bibfnamefont {M.}~\bibnamefont {Zamkov}},\ }\bibfield  {title} {\bibinfo {title} {Quantum {Computing} with {Exciton} {Qubits} in {Colloidal} {Semiconductor} {Nanocrystals}},\ }\href {https://doi.org/10.1021/acs.jpcc.1c05009} {\bibfield  {journal} {\bibinfo  {journal} {J. Phys. Chem. C}\ }\textbf {\bibinfo {volume} {125}},\ \bibinfo {pages} {22195} (\bibinfo {year} {2021})}\BibitemShut {NoStop}%
\bibitem [{\citenamefont {Rubloff}(1972)}]{rubloff_far-ultraviolet_1972}%
  \BibitemOpen
  \bibfield  {author} {\bibinfo {author} {\bibfnamefont {G.~W.}\ \bibnamefont {Rubloff}},\ }\bibfield  {title} {\bibinfo {title} {{Far-Ultraviolet Reflectance Spectra and the Electronic Structure of Ionic Crystals}},\ }\href {https://doi.org/10.1103/PhysRevB.5.662} {\bibfield  {journal} {\bibinfo  {journal} {Phys. Rev. B}\ }\textbf {\bibinfo {volume} {5}},\ \bibinfo {pages} {662} (\bibinfo {year} {1972})}\BibitemShut {NoStop}%
\bibitem [{\citenamefont {Pantelides}(1975)}]{pantelides_electronic_1975}%
  \BibitemOpen
  \bibfield  {author} {\bibinfo {author} {\bibfnamefont {S.~T.}\ \bibnamefont {Pantelides}},\ }\bibfield  {title} {\bibinfo {title} {Electronic excitation energies and the soft-x-ray absorption spectra of alkali halides},\ }\href {https://doi.org/10.1103/PhysRevB.11.2391} {\bibfield  {journal} {\bibinfo  {journal} {Phys. Rev. B}\ }\textbf {\bibinfo {volume} {11}},\ \bibinfo {pages} {2391} (\bibinfo {year} {1975})}\BibitemShut {NoStop}%
\bibitem [{\citenamefont {Haensel}\ \emph {et~al.}(1968)\citenamefont {Haensel}, \citenamefont {Kunz},\ and\ \citenamefont {Sonntag}}]{haensel_measurement_1968}%
  \BibitemOpen
  \bibfield  {author} {\bibinfo {author} {\bibfnamefont {R.}~\bibnamefont {Haensel}}, \bibinfo {author} {\bibfnamefont {C.}~\bibnamefont {Kunz}},\ and\ \bibinfo {author} {\bibfnamefont {B.}~\bibnamefont {Sonntag}},\ }\bibfield  {title} {\bibinfo {title} {{Measurement of Photoabsorption of the Lithium Halides Near the Lithium $K$ Edge}},\ }\href {https://doi.org/10.1103/PhysRevLett.20.262} {\bibfield  {journal} {\bibinfo  {journal} {Phys. Rev. Lett.}\ }\textbf {\bibinfo {volume} {20}},\ \bibinfo {pages} {262} (\bibinfo {year} {1968})}\BibitemShut {NoStop}%
\bibitem [{\citenamefont {Gudat}\ \emph {et~al.}(1974)\citenamefont {Gudat}, \citenamefont {Kunz},\ and\ \citenamefont {Petersen}}]{gudat_core_1974}%
  \BibitemOpen
  \bibfield  {author} {\bibinfo {author} {\bibfnamefont {W.}~\bibnamefont {Gudat}}, \bibinfo {author} {\bibfnamefont {C.}~\bibnamefont {Kunz}},\ and\ \bibinfo {author} {\bibfnamefont {H.}~\bibnamefont {Petersen}},\ }\bibfield  {title} {\bibinfo {title} {{Core Exciton and Band Structure in LiF}},\ }\href {https://doi.org/10.1103/PhysRevLett.32.1370} {\bibfield  {journal} {\bibinfo  {journal} {Phys. Rev. Lett.}\ }\textbf {\bibinfo {volume} {32}},\ \bibinfo {pages} {1370} (\bibinfo {year} {1974})}\BibitemShut {NoStop}%
\bibitem [{\citenamefont {Pantelides}\ and\ \citenamefont {Brown}(1974)}]{pantelides_new_1974}%
  \BibitemOpen
  \bibfield  {author} {\bibinfo {author} {\bibfnamefont {S.~T.}\ \bibnamefont {Pantelides}}\ and\ \bibinfo {author} {\bibfnamefont {F.~C.}\ \bibnamefont {Brown}},\ }\bibfield  {title} {\bibinfo {title} {{New Interpretation of the Soft-X-Ray Absorption Spectra of Several Alkali Halides}},\ }\href {https://doi.org/10.1103/PhysRevLett.33.298} {\bibfield  {journal} {\bibinfo  {journal} {Phys. Rev. Lett.}\ }\textbf {\bibinfo {volume} {33}},\ \bibinfo {pages} {298} (\bibinfo {year} {1974})}\BibitemShut {NoStop}%
\bibitem [{\citenamefont {Kowalczyk}\ \emph {et~al.}(1974)\citenamefont {Kowalczyk}, \citenamefont {McFeely}, \citenamefont {Ley}, \citenamefont {Pollak},\ and\ \citenamefont {Shirley}}]{kowalczyk_x-ray_1974}%
  \BibitemOpen
  \bibfield  {author} {\bibinfo {author} {\bibfnamefont {S.~P.}\ \bibnamefont {Kowalczyk}}, \bibinfo {author} {\bibfnamefont {F.~R.}\ \bibnamefont {McFeely}}, \bibinfo {author} {\bibfnamefont {L.}~\bibnamefont {Ley}}, \bibinfo {author} {\bibfnamefont {R.~A.}\ \bibnamefont {Pollak}},\ and\ \bibinfo {author} {\bibfnamefont {D.~A.}\ \bibnamefont {Shirley}},\ }\bibfield  {title} {\bibinfo {title} {X-ray photoemission studies of the alkali halides},\ }\href {https://doi.org/10.1103/PhysRevB.9.3573} {\bibfield  {journal} {\bibinfo  {journal} {Phys. Rev. B}\ }\textbf {\bibinfo {volume} {9}},\ \bibinfo {pages} {3573} (\bibinfo {year} {1974})}\BibitemShut {NoStop}%
\bibitem [{\citenamefont {Stott}\ \emph {et~al.}(1984)\citenamefont {Stott}, \citenamefont {Hulbert}, \citenamefont {Brown}, \citenamefont {Bunker}, \citenamefont {Chiang}, \citenamefont {Miller},\ and\ \citenamefont {Tan}}]{stott_core_1984}%
  \BibitemOpen
  \bibfield  {author} {\bibinfo {author} {\bibfnamefont {J.~P.}\ \bibnamefont {Stott}}, \bibinfo {author} {\bibfnamefont {S.~L.}\ \bibnamefont {Hulbert}}, \bibinfo {author} {\bibfnamefont {F.~C.}\ \bibnamefont {Brown}}, \bibinfo {author} {\bibfnamefont {B.}~\bibnamefont {Bunker}}, \bibinfo {author} {\bibfnamefont {T.~C.}\ \bibnamefont {Chiang}}, \bibinfo {author} {\bibfnamefont {T.}~\bibnamefont {Miller}},\ and\ \bibinfo {author} {\bibfnamefont {K.~H.}\ \bibnamefont {Tan}},\ }\bibfield  {title} {\bibinfo {title} {{Core excitons at the $K$ edge of LiF}},\ }\href {https://doi.org/10.1103/PhysRevB.30.2163} {\bibfield  {journal} {\bibinfo  {journal} {Phys. Rev. B}\ }\textbf {\bibinfo {volume} {30}},\ \bibinfo {pages} {2163} (\bibinfo {year} {1984})}\BibitemShut {NoStop}%
\bibitem [{\citenamefont {Citrin}(1973)}]{citrin_interatomic_1973}%
  \BibitemOpen
  \bibfield  {author} {\bibinfo {author} {\bibfnamefont {P.~H.}\ \bibnamefont {Citrin}},\ }\bibfield  {title} {\bibinfo {title} {{Interatomic Auger Processes: Effects on Lifetimes of Core Hole States}},\ }\href {https://doi.org/10.1103/PhysRevLett.31.1164} {\bibfield  {journal} {\bibinfo  {journal} {Phys. Rev. Lett.}\ }\textbf {\bibinfo {volume} {31}},\ \bibinfo {pages} {1164} (\bibinfo {year} {1973})}\BibitemShut {NoStop}%
\bibitem [{\citenamefont {Lapeyre}\ \emph {et~al.}(1974)\citenamefont {Lapeyre}, \citenamefont {Baer}, \citenamefont {Hermanson}, \citenamefont {Anderson}, \citenamefont {Knapp},\ and\ \citenamefont {Gobby}}]{lapeyre_photoemission_1974}%
  \BibitemOpen
  \bibfield  {author} {\bibinfo {author} {\bibfnamefont {G.~J.}\ \bibnamefont {Lapeyre}}, \bibinfo {author} {\bibfnamefont {A.~D.}\ \bibnamefont {Baer}}, \bibinfo {author} {\bibfnamefont {J.}~\bibnamefont {Hermanson}}, \bibinfo {author} {\bibfnamefont {J.}~\bibnamefont {Anderson}}, \bibinfo {author} {\bibfnamefont {J.~A.}\ \bibnamefont {Knapp}},\ and\ \bibinfo {author} {\bibfnamefont {P.~L.}\ \bibnamefont {Gobby}},\ }\bibfield  {title} {\bibinfo {title} {Photoemission studies of core exciton decay in {KI}},\ }\href {https://doi.org/10.1016/0038-1098(74)91194-6} {\bibfield  {journal} {\bibinfo  {journal} {Solid State Commun.}\ }\textbf {\bibinfo {volume} {15}},\ \bibinfo {pages} {1601} (\bibinfo {year} {1974})}\BibitemShut {NoStop}%
\bibitem [{\citenamefont {Matthew}\ and\ \citenamefont {Devey}(1974)}]{matthew_breadths_1974}%
  \BibitemOpen
  \bibfield  {author} {\bibinfo {author} {\bibfnamefont {J.~A.~D.}\ \bibnamefont {Matthew}}\ and\ \bibinfo {author} {\bibfnamefont {M.~G.}\ \bibnamefont {Devey}},\ }\bibfield  {title} {\bibinfo {title} {{The breadths of X-ray photoelectron peaks in ionic crystals}},\ }\href {https://doi.org/10.1088/0022-3719/7/17/004} {\bibfield  {journal} {\bibinfo  {journal} {J. Phys. C: Solid State Phys.}\ }\textbf {\bibinfo {volume} {7}},\ \bibinfo {pages} {L335} (\bibinfo {year} {1974})}\BibitemShut {NoStop}%
\bibitem [{\citenamefont {Citrin}\ \emph {et~al.}(1974)\citenamefont {Citrin}, \citenamefont {Eisenberger},\ and\ \citenamefont {Hamann}}]{citrin_phonon_1974}%
  \BibitemOpen
  \bibfield  {author} {\bibinfo {author} {\bibfnamefont {P.~H.}\ \bibnamefont {Citrin}}, \bibinfo {author} {\bibfnamefont {P.}~\bibnamefont {Eisenberger}},\ and\ \bibinfo {author} {\bibfnamefont {D.~R.}\ \bibnamefont {Hamann}},\ }\bibfield  {title} {\bibinfo {title} {{Phonon Broadening of X-Ray Photoemission Linewidths}},\ }\href {https://doi.org/10.1103/PhysRevLett.33.965} {\bibfield  {journal} {\bibinfo  {journal} {Phys. Rev. Lett.}\ }\textbf {\bibinfo {volume} {33}},\ \bibinfo {pages} {965} (\bibinfo {year} {1974})}\BibitemShut {NoStop}%
\bibitem [{\citenamefont {Mahan}(1977)}]{mahan_emission_1977}%
  \BibitemOpen
  \bibfield  {author} {\bibinfo {author} {\bibfnamefont {G.~D.}\ \bibnamefont {Mahan}},\ }\bibfield  {title} {\bibinfo {title} {Emission spectra and phonon relaxation},\ }\href {https://doi.org/10.1103/PhysRevB.15.4587} {\bibfield  {journal} {\bibinfo  {journal} {Phys. Rev. B}\ }\textbf {\bibinfo {volume} {15}},\ \bibinfo {pages} {4587} (\bibinfo {year} {1977})}\BibitemShut {NoStop}%
\bibitem [{\citenamefont {Moulet}\ \emph {et~al.}(2017)\citenamefont {Moulet}, \citenamefont {Bertrand}, \citenamefont {Klostermann}, \citenamefont {Guggenmos}, \citenamefont {Karpowicz},\ and\ \citenamefont {Goulielmakis}}]{moulet_soft_2017}%
  \BibitemOpen
  \bibfield  {author} {\bibinfo {author} {\bibfnamefont {A.}~\bibnamefont {Moulet}}, \bibinfo {author} {\bibfnamefont {J.~B.}\ \bibnamefont {Bertrand}}, \bibinfo {author} {\bibfnamefont {T.}~\bibnamefont {Klostermann}}, \bibinfo {author} {\bibfnamefont {A.}~\bibnamefont {Guggenmos}}, \bibinfo {author} {\bibfnamefont {N.}~\bibnamefont {Karpowicz}},\ and\ \bibinfo {author} {\bibfnamefont {E.}~\bibnamefont {Goulielmakis}},\ }\bibfield  {title} {\bibinfo {title} {Soft x-ray excitonics},\ }\href {https://doi.org/10.1126/science.aan4737} {\bibfield  {journal} {\bibinfo  {journal} {Science}\ }\textbf {\bibinfo {volume} {357}},\ \bibinfo {pages} {1134} (\bibinfo {year} {2017})}\BibitemShut {NoStop}%
\bibitem [{\citenamefont {Chang}\ \emph {et~al.}(2021)\citenamefont {Chang}, \citenamefont {Guggenmos}, \citenamefont {Chen}, \citenamefont {Oh}, \citenamefont {G\'eneaux}, \citenamefont {Chuang}, \citenamefont {Schwartzberg}, \citenamefont {Aloni}, \citenamefont {Neumark},\ and\ \citenamefont {Leone}}]{chang_ws2}%
  \BibitemOpen
  \bibfield  {author} {\bibinfo {author} {\bibfnamefont {H.-T.}\ \bibnamefont {Chang}}, \bibinfo {author} {\bibfnamefont {A.}~\bibnamefont {Guggenmos}}, \bibinfo {author} {\bibfnamefont {C.~T.}\ \bibnamefont {Chen}}, \bibinfo {author} {\bibfnamefont {J.}~\bibnamefont {Oh}}, \bibinfo {author} {\bibfnamefont {R.}~\bibnamefont {G\'eneaux}}, \bibinfo {author} {\bibfnamefont {Y.-D.}\ \bibnamefont {Chuang}}, \bibinfo {author} {\bibfnamefont {A.~M.}\ \bibnamefont {Schwartzberg}}, \bibinfo {author} {\bibfnamefont {S.}~\bibnamefont {Aloni}}, \bibinfo {author} {\bibfnamefont {D.~M.}\ \bibnamefont {Neumark}},\ and\ \bibinfo {author} {\bibfnamefont {S.~R.}\ \bibnamefont {Leone}},\ }\bibfield  {title} {\bibinfo {title} {{Coupled valence carrier and core-exciton dynamics in ${\mathrm{WS}}_{2}$ probed by few-femtosecond extreme ultraviolet transient absorption spectroscopy}},\ }\href {https://doi.org/10.1103/PhysRevB.104.064309} {\bibfield  {journal} {\bibinfo  {journal} {Phys. Rev. B}\ }\textbf {\bibinfo {volume} {104}},\
  \bibinfo {pages} {064309} (\bibinfo {year} {2021})}\BibitemShut {NoStop}%
\bibitem [{\citenamefont {Quintero-Bermudez}\ and\ \citenamefont {Leone}(2024)}]{quintero-bermudez_deciphering_2024}%
  \BibitemOpen
  \bibfield  {author} {\bibinfo {author} {\bibfnamefont {R.}~\bibnamefont {Quintero-Bermudez}}\ and\ \bibinfo {author} {\bibfnamefont {S.~R.}\ \bibnamefont {Leone}},\ }\bibfield  {title} {\bibinfo {title} {{Deciphering core-exciton dynamics in $\mathrm{Ca}{\mathrm{F}}_{2}$ with attosecond spectroscopy}},\ }\href {https://doi.org/10.1103/PhysRevB.109.024308} {\bibfield  {journal} {\bibinfo  {journal} {Phys. Rev. B}\ }\textbf {\bibinfo {volume} {109}},\ \bibinfo {pages} {024308} (\bibinfo {year} {2024})}\BibitemShut {NoStop}%
\bibitem [{\citenamefont {G\'eneaux}\ \emph {et~al.}(2020)\citenamefont {G\'eneaux}, \citenamefont {Kaplan}, \citenamefont {Yue}, \citenamefont {Ross}, \citenamefont {B\ae{}kh\o{}j}, \citenamefont {Kraus}, \citenamefont {Chang}, \citenamefont {Guggenmos}, \citenamefont {Huang}, \citenamefont {Z\"urch}, \citenamefont {Schafer}, \citenamefont {Neumark}, \citenamefont {Gaarde},\ and\ \citenamefont {Leone}}]{geneaux_mgo}%
  \BibitemOpen
  \bibfield  {author} {\bibinfo {author} {\bibfnamefont {R.}~\bibnamefont {G\'eneaux}}, \bibinfo {author} {\bibfnamefont {C.~J.}\ \bibnamefont {Kaplan}}, \bibinfo {author} {\bibfnamefont {L.}~\bibnamefont {Yue}}, \bibinfo {author} {\bibfnamefont {A.~D.}\ \bibnamefont {Ross}}, \bibinfo {author} {\bibfnamefont {J.~E.}\ \bibnamefont {B\ae{}kh\o{}j}}, \bibinfo {author} {\bibfnamefont {P.~M.}\ \bibnamefont {Kraus}}, \bibinfo {author} {\bibfnamefont {H.-T.}\ \bibnamefont {Chang}}, \bibinfo {author} {\bibfnamefont {A.}~\bibnamefont {Guggenmos}}, \bibinfo {author} {\bibfnamefont {M.-Y.}\ \bibnamefont {Huang}}, \bibinfo {author} {\bibfnamefont {M.}~\bibnamefont {Z\"urch}}, \bibinfo {author} {\bibfnamefont {K.~J.}\ \bibnamefont {Schafer}}, \bibinfo {author} {\bibfnamefont {D.~M.}\ \bibnamefont {Neumark}}, \bibinfo {author} {\bibfnamefont {M.~B.}\ \bibnamefont {Gaarde}},\ and\ \bibinfo {author} {\bibfnamefont {S.~R.}\ \bibnamefont {Leone}},\ }\bibfield  {title} {\bibinfo {title} {{Attosecond Time-Domain Measurement of
  Core-Level-Exciton Decay in Magnesium Oxide}},\ }\href {https://doi.org/10.1103/PhysRevLett.124.207401} {\bibfield  {journal} {\bibinfo  {journal} {Phys. Rev. Lett.}\ }\textbf {\bibinfo {volume} {124}},\ \bibinfo {pages} {207401} (\bibinfo {year} {2020})}\BibitemShut {NoStop}%
\bibitem [{\citenamefont {Lucchini}\ \emph {et~al.}(2021)\citenamefont {Lucchini}, \citenamefont {Sato}, \citenamefont {Lucarelli}, \citenamefont {Moio}, \citenamefont {Inzani}, \citenamefont {Borrego-Varillas}, \citenamefont {Frassetto}, \citenamefont {Poletto}, \citenamefont {Hübener}, \citenamefont {De~Giovannini}, \citenamefont {Rubio},\ and\ \citenamefont {Nisoli}}]{lucchini_unravelling_2021}%
  \BibitemOpen
  \bibfield  {author} {\bibinfo {author} {\bibfnamefont {M.}~\bibnamefont {Lucchini}}, \bibinfo {author} {\bibfnamefont {S.~A.}\ \bibnamefont {Sato}}, \bibinfo {author} {\bibfnamefont {G.~D.}\ \bibnamefont {Lucarelli}}, \bibinfo {author} {\bibfnamefont {B.}~\bibnamefont {Moio}}, \bibinfo {author} {\bibfnamefont {G.}~\bibnamefont {Inzani}}, \bibinfo {author} {\bibfnamefont {R.}~\bibnamefont {Borrego-Varillas}}, \bibinfo {author} {\bibfnamefont {F.}~\bibnamefont {Frassetto}}, \bibinfo {author} {\bibfnamefont {L.}~\bibnamefont {Poletto}}, \bibinfo {author} {\bibfnamefont {H.}~\bibnamefont {Hübener}}, \bibinfo {author} {\bibfnamefont {U.}~\bibnamefont {De~Giovannini}}, \bibinfo {author} {\bibfnamefont {A.}~\bibnamefont {Rubio}},\ and\ \bibinfo {author} {\bibfnamefont {M.}~\bibnamefont {Nisoli}},\ }\bibfield  {title} {\bibinfo {title} {Unravelling the intertwined atomic and bulk nature of localised excitons by attosecond spectroscopy},\ }\href {https://doi.org/10.1038/s41467-021-21345-7} {\bibfield  {journal}
  {\bibinfo  {journal} {Nat. Commun.}\ }\textbf {\bibinfo {volume} {12}},\ \bibinfo {pages} {1021} (\bibinfo {year} {2021})}\BibitemShut {NoStop}%
\bibitem [{\citenamefont {Gaynor}\ \emph {et~al.}(2021)\citenamefont {Gaynor}, \citenamefont {Fidler}, \citenamefont {Lin}, \citenamefont {Chang}, \citenamefont {Zuerch}, \citenamefont {Neumark},\ and\ \citenamefont {Leone}}]{gaynor_nacl}%
  \BibitemOpen
  \bibfield  {author} {\bibinfo {author} {\bibfnamefont {J.~D.}\ \bibnamefont {Gaynor}}, \bibinfo {author} {\bibfnamefont {A.~P.}\ \bibnamefont {Fidler}}, \bibinfo {author} {\bibfnamefont {Y.-C.}\ \bibnamefont {Lin}}, \bibinfo {author} {\bibfnamefont {H.-T.}\ \bibnamefont {Chang}}, \bibinfo {author} {\bibfnamefont {M.}~\bibnamefont {Zuerch}}, \bibinfo {author} {\bibfnamefont {D.~M.}\ \bibnamefont {Neumark}},\ and\ \bibinfo {author} {\bibfnamefont {S.~R.}\ \bibnamefont {Leone}},\ }\bibfield  {title} {\bibinfo {title} {{Solid state core-exciton dynamics in NaCl observed by tabletop attosecond four-wave mixing spectroscopy}},\ }\href {https://doi.org/10.1103/PhysRevB.103.245140} {\bibfield  {journal} {\bibinfo  {journal} {Phys. Rev. B}\ }\textbf {\bibinfo {volume} {103}},\ \bibinfo {pages} {245140} (\bibinfo {year} {2021})}\BibitemShut {NoStop}%
\bibitem [{\citenamefont {Moore}(1949)}]{moore_atomic_1949}%
  \BibitemOpen
  \bibfield  {author} {\bibinfo {author} {\bibfnamefont {C.~E.}\ \bibnamefont {Moore}},\ }\href {https://apps.dtic.mil/sti/citations/tr/ADA278130} {\emph {\bibinfo {title} {{Atomic Energy Levels}}}},\ \bibinfo {series} {Circulation}, Vol.~\bibinfo {volume} {I}\ (\bibinfo  {publisher} {National Bureau of Standards},\ \bibinfo {address} {U.S. GPO, Washington, D.C.},\ \bibinfo {year} {1949})\BibitemShut {NoStop}%
\bibitem [{\citenamefont {Kunz}\ \emph {et~al.}(1973)\citenamefont {Kunz}, \citenamefont {Mickish},\ and\ \citenamefont {Collins}}]{kunz_absorption_1973}%
  \BibitemOpen
  \bibfield  {author} {\bibinfo {author} {\bibfnamefont {A.~B.}\ \bibnamefont {Kunz}}, \bibinfo {author} {\bibfnamefont {D.~J.}\ \bibnamefont {Mickish}},\ and\ \bibinfo {author} {\bibfnamefont {T.~C.}\ \bibnamefont {Collins}},\ }\bibfield  {title} {\bibinfo {title} {{Absorption of Soft X Rays by Insulators with a Forbidden Exciton Transition}},\ }\href {https://doi.org/10.1103/PhysRevLett.31.756} {\bibfield  {journal} {\bibinfo  {journal} {Phys. Rev. Lett.}\ }\textbf {\bibinfo {volume} {31}},\ \bibinfo {pages} {756} (\bibinfo {year} {1973})}\BibitemShut {NoStop}%
\bibitem [{\citenamefont {Sonntag}(1974)}]{sonntag_observations_1974}%
  \BibitemOpen
  \bibfield  {author} {\bibinfo {author} {\bibfnamefont {B.~F.}\ \bibnamefont {Sonntag}},\ }\bibfield  {title} {\bibinfo {title} {{Observations of "forbidden" soft-x-ray transitions: Li $\mathrm{} K$ absorption in LiF}},\ }\href {https://doi.org/10.1103/PhysRevB.9.3601} {\bibfield  {journal} {\bibinfo  {journal} {Phys. Rev. B}\ }\textbf {\bibinfo {volume} {9}},\ \bibinfo {pages} {3601} (\bibinfo {year} {1974})}\BibitemShut {NoStop}%
\bibitem [{\citenamefont {Fields}\ \emph {et~al.}(1977)\citenamefont {Fields}, \citenamefont {Gibbons},\ and\ \citenamefont {Schnatterly}}]{fields_electronic_1977}%
  \BibitemOpen
  \bibfield  {author} {\bibinfo {author} {\bibfnamefont {J.~R.}\ \bibnamefont {Fields}}, \bibinfo {author} {\bibfnamefont {P.~C.}\ \bibnamefont {Gibbons}},\ and\ \bibinfo {author} {\bibfnamefont {S.~E.}\ \bibnamefont {Schnatterly}},\ }\bibfield  {title} {\bibinfo {title} {{Electronic Excitations in LiF: 10-70 eV}},\ }\href {https://doi.org/10.1103/PhysRevLett.38.430} {\bibfield  {journal} {\bibinfo  {journal} {Phys. Rev. Lett.}\ }\textbf {\bibinfo {volume} {38}},\ \bibinfo {pages} {430} (\bibinfo {year} {1977})}\BibitemShut {NoStop}%
\bibitem [{\citenamefont {Olovsson}\ \emph {et~al.}(2009)\citenamefont {Olovsson}, \citenamefont {Tanaka}, \citenamefont {Puschnig},\ and\ \citenamefont {Ambrosch-Draxl}}]{olovsson_near-edge_2009}%
  \BibitemOpen
  \bibfield  {author} {\bibinfo {author} {\bibfnamefont {W.}~\bibnamefont {Olovsson}}, \bibinfo {author} {\bibfnamefont {I.}~\bibnamefont {Tanaka}}, \bibinfo {author} {\bibfnamefont {P.}~\bibnamefont {Puschnig}},\ and\ \bibinfo {author} {\bibfnamefont {C.}~\bibnamefont {Ambrosch-Draxl}},\ }\bibfield  {title} {\bibinfo {title} {{Near-edge structures from first principles all-electron Bethe–Salpeter equation calculations}},\ }\href {https://doi.org/10.1088/0953-8984/21/10/104205} {\bibfield  {journal} {\bibinfo  {journal} {J. Phys.: Condens. Matter}\ }\textbf {\bibinfo {volume} {21}},\ \bibinfo {pages} {104205} (\bibinfo {year} {2009})}\BibitemShut {NoStop}%
\bibitem [{\citenamefont {Kunz}\ \emph {et~al.}(1972)\citenamefont {Kunz}, \citenamefont {Devreese},\ and\ \citenamefont {Collins}}]{kunz_role_1972}%
  \BibitemOpen
  \bibfield  {author} {\bibinfo {author} {\bibfnamefont {A.~B.}\ \bibnamefont {Kunz}}, \bibinfo {author} {\bibfnamefont {J.~T.}\ \bibnamefont {Devreese}},\ and\ \bibinfo {author} {\bibfnamefont {T.}~\bibnamefont {Collins}},\ }\bibfield  {title} {\bibinfo {title} {The role of the electronic polaron in the soft {X}-ray absorption of the lithium halides},\ }\href {https://doi.org/10.1088/0022-3719/5/22/012} {\bibfield  {journal} {\bibinfo  {journal} {J. Phys. C: Solid State Phys.}\ }\textbf {\bibinfo {volume} {5}},\ \bibinfo {pages} {3259} (\bibinfo {year} {1972})}\BibitemShut {NoStop}%
\bibitem [{\citenamefont {Rottke}\ \emph {et~al.}(2022)\citenamefont {Rottke}, \citenamefont {Engel}, \citenamefont {Schick}, \citenamefont {Schunck}, \citenamefont {Miedema}, \citenamefont {Borchert}, \citenamefont {Kuhlmann}, \citenamefont {Ekanayake}, \citenamefont {Dziarzhytski}, \citenamefont {Brenner}, \citenamefont {Eichmann}, \citenamefont {von Korff~Schmising}, \citenamefont {Beye},\ and\ \citenamefont {Eisebitt}}]{rottke_probing_2022}%
  \BibitemOpen
  \bibfield  {author} {\bibinfo {author} {\bibfnamefont {H.}~\bibnamefont {Rottke}}, \bibinfo {author} {\bibfnamefont {R.~Y.}\ \bibnamefont {Engel}}, \bibinfo {author} {\bibfnamefont {D.}~\bibnamefont {Schick}}, \bibinfo {author} {\bibfnamefont {J.~O.}\ \bibnamefont {Schunck}}, \bibinfo {author} {\bibfnamefont {P.~S.}\ \bibnamefont {Miedema}}, \bibinfo {author} {\bibfnamefont {M.~C.}\ \bibnamefont {Borchert}}, \bibinfo {author} {\bibfnamefont {M.}~\bibnamefont {Kuhlmann}}, \bibinfo {author} {\bibfnamefont {N.}~\bibnamefont {Ekanayake}}, \bibinfo {author} {\bibfnamefont {S.}~\bibnamefont {Dziarzhytski}}, \bibinfo {author} {\bibfnamefont {G.}~\bibnamefont {Brenner}}, \bibinfo {author} {\bibfnamefont {U.}~\bibnamefont {Eichmann}}, \bibinfo {author} {\bibfnamefont {C.}~\bibnamefont {von Korff~Schmising}}, \bibinfo {author} {\bibfnamefont {M.}~\bibnamefont {Beye}},\ and\ \bibinfo {author} {\bibfnamefont {S.}~\bibnamefont {Eisebitt}},\ }\bibfield  {title} {\bibinfo {title} {Probing electron and hole colocalization
  by resonant four-wave mixing spectroscopy in the extreme ultraviolet},\ }\href {https://doi.org/10.1126/sciadv.abn5127} {\bibfield  {journal} {\bibinfo  {journal} {Sci. Adv.}\ }\textbf {\bibinfo {volume} {8}},\ \bibinfo {pages} {eabn5127} (\bibinfo {year} {2022})}\BibitemShut {NoStop}%
\bibitem [{\citenamefont {Zinchenko}\ \emph {et~al.}(2023)\citenamefont {Zinchenko}, \citenamefont {Ardana-Lamas}, \citenamefont {Lanfaloni}, \citenamefont {Luu}, \citenamefont {Pertot}, \citenamefont {Huppert},\ and\ \citenamefont {Wörner}}]{zinchenko_apparatus_2023}%
  \BibitemOpen
  \bibfield  {author} {\bibinfo {author} {\bibfnamefont {K.~S.}\ \bibnamefont {Zinchenko}}, \bibinfo {author} {\bibfnamefont {F.}~\bibnamefont {Ardana-Lamas}}, \bibinfo {author} {\bibfnamefont {V.~U.}\ \bibnamefont {Lanfaloni}}, \bibinfo {author} {\bibfnamefont {T.~T.}\ \bibnamefont {Luu}}, \bibinfo {author} {\bibfnamefont {Y.}~\bibnamefont {Pertot}}, \bibinfo {author} {\bibfnamefont {M.}~\bibnamefont {Huppert}},\ and\ \bibinfo {author} {\bibfnamefont {H.~J.}\ \bibnamefont {Wörner}},\ }\bibfield  {title} {\bibinfo {title} {{Apparatus for attosecond transient-absorption spectroscopy in the water-window soft-X-ray region}},\ }\href {https://doi.org/10.1038/s41598-023-29089-8} {\bibfield  {journal} {\bibinfo  {journal} {Sci. Rep.}\ }\textbf {\bibinfo {volume} {13}},\ \bibinfo {pages} {3059} (\bibinfo {year} {2023})}\BibitemShut {NoStop}%
\bibitem [{\citenamefont {Chen}\ \emph {et~al.}(2013)\citenamefont {Chen}, \citenamefont {Wu}, \citenamefont {Gaarde},\ and\ \citenamefont {Schafer}}]{chen_quantum_2013}%
  \BibitemOpen
  \bibfield  {author} {\bibinfo {author} {\bibfnamefont {S.}~\bibnamefont {Chen}}, \bibinfo {author} {\bibfnamefont {M.}~\bibnamefont {Wu}}, \bibinfo {author} {\bibfnamefont {M.~B.}\ \bibnamefont {Gaarde}},\ and\ \bibinfo {author} {\bibfnamefont {K.~J.}\ \bibnamefont {Schafer}},\ }\bibfield  {title} {\bibinfo {title} {Quantum interference in attosecond transient absorption of laser-dressed helium atoms},\ }\href {https://doi.org/10.1103/PhysRevA.87.033408} {\bibfield  {journal} {\bibinfo  {journal} {Phys. Rev. A}\ }\textbf {\bibinfo {volume} {87}},\ \bibinfo {pages} {033408} (\bibinfo {year} {2013})}\BibitemShut {NoStop}%
\bibitem [{\citenamefont {Lucchini}\ \emph {et~al.}(2016)\citenamefont {Lucchini}, \citenamefont {Sato}, \citenamefont {Ludwig}, \citenamefont {Herrmann}, \citenamefont {Volkov}, \citenamefont {Kasmi}, \citenamefont {Shinohara}, \citenamefont {Yabana}, \citenamefont {Gallmann},\ and\ \citenamefont {Keller}}]{lucchini_attosecond_2016}%
  \BibitemOpen
  \bibfield  {author} {\bibinfo {author} {\bibfnamefont {M.}~\bibnamefont {Lucchini}}, \bibinfo {author} {\bibfnamefont {S.~A.}\ \bibnamefont {Sato}}, \bibinfo {author} {\bibfnamefont {A.}~\bibnamefont {Ludwig}}, \bibinfo {author} {\bibfnamefont {J.}~\bibnamefont {Herrmann}}, \bibinfo {author} {\bibfnamefont {M.}~\bibnamefont {Volkov}}, \bibinfo {author} {\bibfnamefont {L.}~\bibnamefont {Kasmi}}, \bibinfo {author} {\bibfnamefont {Y.}~\bibnamefont {Shinohara}}, \bibinfo {author} {\bibfnamefont {K.}~\bibnamefont {Yabana}}, \bibinfo {author} {\bibfnamefont {L.}~\bibnamefont {Gallmann}},\ and\ \bibinfo {author} {\bibfnamefont {U.}~\bibnamefont {Keller}},\ }\bibfield  {title} {\bibinfo {title} {{Attosecond dynamical Franz-Keldysh effect in polycrystalline diamond}},\ }\href {https://doi.org/10.1126/science.aag1268} {\bibfield  {journal} {\bibinfo  {journal} {Science}\ }\textbf {\bibinfo {volume} {353}},\ \bibinfo {pages} {916} (\bibinfo {year} {2016})}\BibitemShut {NoStop}%
\bibitem [{\citenamefont {Combescot}(1992)}]{combescot_semiconductors_1992}%
  \BibitemOpen
  \bibfield  {author} {\bibinfo {author} {\bibfnamefont {M.}~\bibnamefont {Combescot}},\ }\bibfield  {title} {\bibinfo {title} {{Semiconductors in strong laser fields: from polariton to exciton optical Stark effect}},\ }\href {https://doi.org/10.1016/0370-1573(92)90103-7} {\bibfield  {journal} {\bibinfo  {journal} {Phys. Rep.}\ }\textbf {\bibinfo {volume} {221}},\ \bibinfo {pages} {167} (\bibinfo {year} {1992})}\BibitemShut {NoStop}%
\bibitem [{\citenamefont {Wu}\ \emph {et~al.}(2016)\citenamefont {Wu}, \citenamefont {Chen}, \citenamefont {Camp}, \citenamefont {Schafer},\ and\ \citenamefont {Gaarde}}]{wu_theory_2016}%
  \BibitemOpen
  \bibfield  {author} {\bibinfo {author} {\bibfnamefont {M.}~\bibnamefont {Wu}}, \bibinfo {author} {\bibfnamefont {S.}~\bibnamefont {Chen}}, \bibinfo {author} {\bibfnamefont {S.}~\bibnamefont {Camp}}, \bibinfo {author} {\bibfnamefont {K.~J.}\ \bibnamefont {Schafer}},\ and\ \bibinfo {author} {\bibfnamefont {M.~B.}\ \bibnamefont {Gaarde}},\ }\bibfield  {title} {\bibinfo {title} {Theory of strong-field attosecond transient absorption},\ }\href {https://doi.org/10.1088/0953-4075/49/6/062003} {\bibfield  {journal} {\bibinfo  {journal} {J. Phys. B: At. Mol. Opt. Phys.}\ }\textbf {\bibinfo {volume} {49}},\ \bibinfo {pages} {062003} (\bibinfo {year} {2016})},\ \bibinfo {note} {publisher: IOP Publishing}\BibitemShut {NoStop}%
\bibitem [{\citenamefont {Citrin}\ \emph {et~al.}(1977)\citenamefont {Citrin}, \citenamefont {Wertheim},\ and\ \citenamefont {Baer}}]{citrin_many-body_1977}%
  \BibitemOpen
  \bibfield  {author} {\bibinfo {author} {\bibfnamefont {P.~H.}\ \bibnamefont {Citrin}}, \bibinfo {author} {\bibfnamefont {G.~K.}\ \bibnamefont {Wertheim}},\ and\ \bibinfo {author} {\bibfnamefont {Y.}~\bibnamefont {Baer}},\ }\bibfield  {title} {\bibinfo {title} {{Many-body processes in x-ray photoemission line shapes from Li, Na, Mg, and Al metals}},\ }\href {https://doi.org/10.1103/PhysRevB.16.4256} {\bibfield  {journal} {\bibinfo  {journal} {Phys. Rev. B}\ }\textbf {\bibinfo {volume} {16}},\ \bibinfo {pages} {4256} (\bibinfo {year} {1977})}\BibitemShut {NoStop}%
\bibitem [{\citenamefont {Giannozzi}\ \emph {et~al.}(2009)\citenamefont {Giannozzi}, \citenamefont {Baroni}, \citenamefont {Bonini}, \citenamefont {Calandra}, \citenamefont {Car}, \citenamefont {Cavazzoni}, \citenamefont {Ceresoli}, \citenamefont {Chiarotti}, \citenamefont {Cococcioni}, \citenamefont {Dabo}, \citenamefont {Corso}, \citenamefont {Gironcoli}, \citenamefont {Fabris}, \citenamefont {Fratesi}, \citenamefont {Gebauer}, \citenamefont {Gerstmann}, \citenamefont {Gougoussis}, \citenamefont {Kokalj}, \citenamefont {Lazzeri}, \citenamefont {Martin-Samos}, \citenamefont {Marzari}, \citenamefont {Mauri}, \citenamefont {Mazzarello}, \citenamefont {Paolini}, \citenamefont {Pasquarello}, \citenamefont {Paulatto}, \citenamefont {Sbraccia}, \citenamefont {Scandolo}, \citenamefont {Sclauzero}, \citenamefont {Seitsonen}, \citenamefont {Smogunov}, \citenamefont {Umari},\ and\ \citenamefont {Wentzcovitch}}]{giannozzi_quantum_2009}%
  \BibitemOpen
  \bibfield  {author} {\bibinfo {author} {\bibfnamefont {P.}~\bibnamefont {Giannozzi}}, \bibinfo {author} {\bibfnamefont {S.}~\bibnamefont {Baroni}}, \bibinfo {author} {\bibfnamefont {N.}~\bibnamefont {Bonini}}, \bibinfo {author} {\bibfnamefont {M.}~\bibnamefont {Calandra}}, \bibinfo {author} {\bibfnamefont {R.}~\bibnamefont {Car}}, \bibinfo {author} {\bibfnamefont {C.}~\bibnamefont {Cavazzoni}}, \bibinfo {author} {\bibfnamefont {D.}~\bibnamefont {Ceresoli}}, \bibinfo {author} {\bibfnamefont {G.~L.}\ \bibnamefont {Chiarotti}}, \bibinfo {author} {\bibfnamefont {M.}~\bibnamefont {Cococcioni}}, \bibinfo {author} {\bibfnamefont {I.}~\bibnamefont {Dabo}}, \bibinfo {author} {\bibfnamefont {A.~D.}\ \bibnamefont {Corso}}, \bibinfo {author} {\bibfnamefont {S.~d.}\ \bibnamefont {Gironcoli}}, \bibinfo {author} {\bibfnamefont {S.}~\bibnamefont {Fabris}}, \bibinfo {author} {\bibfnamefont {G.}~\bibnamefont {Fratesi}}, \bibinfo {author} {\bibfnamefont {R.}~\bibnamefont {Gebauer}}, \bibinfo {author} {\bibfnamefont
  {U.}~\bibnamefont {Gerstmann}}, \bibinfo {author} {\bibfnamefont {C.}~\bibnamefont {Gougoussis}}, \bibinfo {author} {\bibfnamefont {A.}~\bibnamefont {Kokalj}}, \bibinfo {author} {\bibfnamefont {M.}~\bibnamefont {Lazzeri}}, \bibinfo {author} {\bibfnamefont {L.}~\bibnamefont {Martin-Samos}}, \bibinfo {author} {\bibfnamefont {N.}~\bibnamefont {Marzari}}, \bibinfo {author} {\bibfnamefont {F.}~\bibnamefont {Mauri}}, \bibinfo {author} {\bibfnamefont {R.}~\bibnamefont {Mazzarello}}, \bibinfo {author} {\bibfnamefont {S.}~\bibnamefont {Paolini}}, \bibinfo {author} {\bibfnamefont {A.}~\bibnamefont {Pasquarello}}, \bibinfo {author} {\bibfnamefont {L.}~\bibnamefont {Paulatto}}, \bibinfo {author} {\bibfnamefont {C.}~\bibnamefont {Sbraccia}}, \bibinfo {author} {\bibfnamefont {S.}~\bibnamefont {Scandolo}}, \bibinfo {author} {\bibfnamefont {G.}~\bibnamefont {Sclauzero}}, \bibinfo {author} {\bibfnamefont {A.~P.}\ \bibnamefont {Seitsonen}}, \bibinfo {author} {\bibfnamefont {A.}~\bibnamefont {Smogunov}}, \bibinfo {author}
  {\bibfnamefont {P.}~\bibnamefont {Umari}},\ and\ \bibinfo {author} {\bibfnamefont {R.~M.}\ \bibnamefont {Wentzcovitch}},\ }\bibfield  {title} {\bibinfo {title} {{QUANTUM} {ESPRESSO}: a modular and open-source software project for quantum simulations of materials},\ }\href {https://doi.org/10.1088/0953-8984/21/39/395502} {\bibfield  {journal} {\bibinfo  {journal} {J. Phys.: Condens. Matter}\ }\textbf {\bibinfo {volume} {21}},\ \bibinfo {pages} {395502} (\bibinfo {year} {2009})}\BibitemShut {NoStop}%
\bibitem [{\citenamefont {Giannozzi}\ \emph {et~al.}(2017)\citenamefont {Giannozzi}, \citenamefont {Andreussi}, \citenamefont {Brumme}, \citenamefont {Bunau}, \citenamefont {Nardelli}, \citenamefont {Calandra}, \citenamefont {Car}, \citenamefont {Cavazzoni}, \citenamefont {Ceresoli}, \citenamefont {Cococcioni}, \citenamefont {Colonna}, \citenamefont {Carnimeo}, \citenamefont {Corso}, \citenamefont {Gironcoli}, \citenamefont {Delugas}, \citenamefont {DiStasio}, \citenamefont {Ferretti}, \citenamefont {Floris}, \citenamefont {Fratesi}, \citenamefont {Fugallo}, \citenamefont {Gebauer}, \citenamefont {Gerstmann}, \citenamefont {Giustino}, \citenamefont {Gorni}, \citenamefont {Jia}, \citenamefont {Kawamura}, \citenamefont {Ko}, \citenamefont {Kokalj}, \citenamefont {Küçükbenli}, \citenamefont {Lazzeri}, \citenamefont {Marsili}, \citenamefont {Marzari}, \citenamefont {Mauri}, \citenamefont {Nguyen}, \citenamefont {Nguyen}, \citenamefont {Otero-de-la Roza}, \citenamefont {Paulatto}, \citenamefont {Poncé},
  \citenamefont {Rocca}, \citenamefont {Sabatini}, \citenamefont {Santra}, \citenamefont {Schlipf}, \citenamefont {Seitsonen}, \citenamefont {Smogunov}, \citenamefont {Timrov}, \citenamefont {Thonhauser}, \citenamefont {Umari}, \citenamefont {Vast}, \citenamefont {Wu},\ and\ \citenamefont {Baroni}}]{giannozzi_advanced_2017}%
  \BibitemOpen
  \bibfield  {author} {\bibinfo {author} {\bibfnamefont {P.}~\bibnamefont {Giannozzi}}, \bibinfo {author} {\bibfnamefont {O.}~\bibnamefont {Andreussi}}, \bibinfo {author} {\bibfnamefont {T.}~\bibnamefont {Brumme}}, \bibinfo {author} {\bibfnamefont {O.}~\bibnamefont {Bunau}}, \bibinfo {author} {\bibfnamefont {M.~B.}\ \bibnamefont {Nardelli}}, \bibinfo {author} {\bibfnamefont {M.}~\bibnamefont {Calandra}}, \bibinfo {author} {\bibfnamefont {R.}~\bibnamefont {Car}}, \bibinfo {author} {\bibfnamefont {C.}~\bibnamefont {Cavazzoni}}, \bibinfo {author} {\bibfnamefont {D.}~\bibnamefont {Ceresoli}}, \bibinfo {author} {\bibfnamefont {M.}~\bibnamefont {Cococcioni}}, \bibinfo {author} {\bibfnamefont {N.}~\bibnamefont {Colonna}}, \bibinfo {author} {\bibfnamefont {I.}~\bibnamefont {Carnimeo}}, \bibinfo {author} {\bibfnamefont {A.~D.}\ \bibnamefont {Corso}}, \bibinfo {author} {\bibfnamefont {S.~d.}\ \bibnamefont {Gironcoli}}, \bibinfo {author} {\bibfnamefont {P.}~\bibnamefont {Delugas}}, \bibinfo {author} {\bibfnamefont
  {R.~A.}\ \bibnamefont {DiStasio}}, \bibinfo {author} {\bibfnamefont {A.}~\bibnamefont {Ferretti}}, \bibinfo {author} {\bibfnamefont {A.}~\bibnamefont {Floris}}, \bibinfo {author} {\bibfnamefont {G.}~\bibnamefont {Fratesi}}, \bibinfo {author} {\bibfnamefont {G.}~\bibnamefont {Fugallo}}, \bibinfo {author} {\bibfnamefont {R.}~\bibnamefont {Gebauer}}, \bibinfo {author} {\bibfnamefont {U.}~\bibnamefont {Gerstmann}}, \bibinfo {author} {\bibfnamefont {F.}~\bibnamefont {Giustino}}, \bibinfo {author} {\bibfnamefont {T.}~\bibnamefont {Gorni}}, \bibinfo {author} {\bibfnamefont {J.}~\bibnamefont {Jia}}, \bibinfo {author} {\bibfnamefont {M.}~\bibnamefont {Kawamura}}, \bibinfo {author} {\bibfnamefont {H.-Y.}\ \bibnamefont {Ko}}, \bibinfo {author} {\bibfnamefont {A.}~\bibnamefont {Kokalj}}, \bibinfo {author} {\bibfnamefont {E.}~\bibnamefont {Küçükbenli}}, \bibinfo {author} {\bibfnamefont {M.}~\bibnamefont {Lazzeri}}, \bibinfo {author} {\bibfnamefont {M.}~\bibnamefont {Marsili}}, \bibinfo {author} {\bibfnamefont
  {N.}~\bibnamefont {Marzari}}, \bibinfo {author} {\bibfnamefont {F.}~\bibnamefont {Mauri}}, \bibinfo {author} {\bibfnamefont {N.~L.}\ \bibnamefont {Nguyen}}, \bibinfo {author} {\bibfnamefont {H.-V.}\ \bibnamefont {Nguyen}}, \bibinfo {author} {\bibfnamefont {A.}~\bibnamefont {Otero-de-la Roza}}, \bibinfo {author} {\bibfnamefont {L.}~\bibnamefont {Paulatto}}, \bibinfo {author} {\bibfnamefont {S.}~\bibnamefont {Poncé}}, \bibinfo {author} {\bibfnamefont {D.}~\bibnamefont {Rocca}}, \bibinfo {author} {\bibfnamefont {R.}~\bibnamefont {Sabatini}}, \bibinfo {author} {\bibfnamefont {B.}~\bibnamefont {Santra}}, \bibinfo {author} {\bibfnamefont {M.}~\bibnamefont {Schlipf}}, \bibinfo {author} {\bibfnamefont {A.~P.}\ \bibnamefont {Seitsonen}}, \bibinfo {author} {\bibfnamefont {A.}~\bibnamefont {Smogunov}}, \bibinfo {author} {\bibfnamefont {I.}~\bibnamefont {Timrov}}, \bibinfo {author} {\bibfnamefont {T.}~\bibnamefont {Thonhauser}}, \bibinfo {author} {\bibfnamefont {P.}~\bibnamefont {Umari}}, \bibinfo {author}
  {\bibfnamefont {N.}~\bibnamefont {Vast}}, \bibinfo {author} {\bibfnamefont {X.}~\bibnamefont {Wu}},\ and\ \bibinfo {author} {\bibfnamefont {S.}~\bibnamefont {Baroni}},\ }\bibfield  {title} {\bibinfo {title} {Advanced capabilities for materials modelling with {Quantum} {ESPRESSO}},\ }\href {https://doi.org/10.1088/1361-648X/aa8f79} {\bibfield  {journal} {\bibinfo  {journal} {J. Phys.: Condens. Matter}\ }\textbf {\bibinfo {volume} {29}},\ \bibinfo {pages} {465901} (\bibinfo {year} {2017})}\BibitemShut {NoStop}%
\bibitem [{\citenamefont {Wang}\ \emph {et~al.}(2003)\citenamefont {Wang}, \citenamefont {Rohlfing}, \citenamefont {Kr\"uger},\ and\ \citenamefont {Pollmann}}]{wang_quasiparticle_2003}%
  \BibitemOpen
  \bibfield  {author} {\bibinfo {author} {\bibfnamefont {N.-P.}\ \bibnamefont {Wang}}, \bibinfo {author} {\bibfnamefont {M.}~\bibnamefont {Rohlfing}}, \bibinfo {author} {\bibfnamefont {P.}~\bibnamefont {Kr\"uger}},\ and\ \bibinfo {author} {\bibfnamefont {J.}~\bibnamefont {Pollmann}},\ }\bibfield  {title} {\bibinfo {title} {{Quasiparticle band structure and optical spectrum of LiF(001)}},\ }\href {https://doi.org/10.1103/PhysRevB.67.115111} {\bibfield  {journal} {\bibinfo  {journal} {Phys. Rev. B}\ }\textbf {\bibinfo {volume} {67}},\ \bibinfo {pages} {115111} (\bibinfo {year} {2003})}\BibitemShut {NoStop}%
\bibitem [{\citenamefont {Sommer}\ \emph {et~al.}(2012)\citenamefont {Sommer}, \citenamefont {Kr\"uger},\ and\ \citenamefont {Pollmann}}]{sommer_optical_2012}%
  \BibitemOpen
  \bibfield  {author} {\bibinfo {author} {\bibfnamefont {C.}~\bibnamefont {Sommer}}, \bibinfo {author} {\bibfnamefont {P.}~\bibnamefont {Kr\"uger}},\ and\ \bibinfo {author} {\bibfnamefont {J.}~\bibnamefont {Pollmann}},\ }\bibfield  {title} {\bibinfo {title} {Optical spectra of alkali-metal fluorides},\ }\href {https://doi.org/10.1103/PhysRevB.86.155212} {\bibfield  {journal} {\bibinfo  {journal} {Phys. Rev. B}\ }\textbf {\bibinfo {volume} {86}},\ \bibinfo {pages} {155212} (\bibinfo {year} {2012})}\BibitemShut {NoStop}%
\bibitem [{\citenamefont {Pascal}\ \emph {et~al.}(2014)\citenamefont {Pascal}, \citenamefont {Boesenberg}, \citenamefont {Kostecki}, \citenamefont {Richardson}, \citenamefont {Weng}, \citenamefont {Sokaras}, \citenamefont {Nordlund}, \citenamefont {McDermott}, \citenamefont {Moewes}, \citenamefont {Cabana},\ and\ \citenamefont {Prendergast}}]{pascal_finite_2014}%
  \BibitemOpen
  \bibfield  {author} {\bibinfo {author} {\bibfnamefont {T.~A.}\ \bibnamefont {Pascal}}, \bibinfo {author} {\bibfnamefont {U.}~\bibnamefont {Boesenberg}}, \bibinfo {author} {\bibfnamefont {R.}~\bibnamefont {Kostecki}}, \bibinfo {author} {\bibfnamefont {T.~J.}\ \bibnamefont {Richardson}}, \bibinfo {author} {\bibfnamefont {T.-C.}\ \bibnamefont {Weng}}, \bibinfo {author} {\bibfnamefont {D.}~\bibnamefont {Sokaras}}, \bibinfo {author} {\bibfnamefont {D.}~\bibnamefont {Nordlund}}, \bibinfo {author} {\bibfnamefont {E.}~\bibnamefont {McDermott}}, \bibinfo {author} {\bibfnamefont {A.}~\bibnamefont {Moewes}}, \bibinfo {author} {\bibfnamefont {J.}~\bibnamefont {Cabana}},\ and\ \bibinfo {author} {\bibfnamefont {D.}~\bibnamefont {Prendergast}},\ }\bibfield  {title} {\bibinfo {title} {Finite temperature effects on the {X}-ray absorption spectra of lithium compounds: {First}-principles interpretation of {X}-ray {Raman} measurements},\ }\href {https://doi.org/10.1063/1.4856835} {\bibfield  {journal} {\bibinfo  {journal} {J.
  Chem. Phys.}\ }\textbf {\bibinfo {volume} {140}},\ \bibinfo {pages} {034107} (\bibinfo {year} {2014})}\BibitemShut {NoStop}%
\bibitem [{\citenamefont {Rajput}\ \emph {et~al.}(2022)\citenamefont {Rajput}, \citenamefont {Kumar},\ and\ \citenamefont {Roy}}]{rajput_two-dimensional_2022}%
  \BibitemOpen
  \bibfield  {author} {\bibinfo {author} {\bibfnamefont {K.}~\bibnamefont {Rajput}}, \bibinfo {author} {\bibfnamefont {V.}~\bibnamefont {Kumar}},\ and\ \bibinfo {author} {\bibfnamefont {D.~R.}\ \bibnamefont {Roy}},\ }\bibfield  {title} {\bibinfo {title} {Two-{Dimensional} lithium fluoride ({LiF}) as an efficient hydrogen storage material},\ }\href {https://doi.org/10.1016/j.apsusc.2021.151776} {\bibfield  {journal} {\bibinfo  {journal} {Appl. Surf. Sci.}\ }\textbf {\bibinfo {volume} {581}},\ \bibinfo {pages} {151776} (\bibinfo {year} {2022})}\BibitemShut {NoStop}%
\bibitem [{\citenamefont {Gulans}\ \emph {et~al.}(2014)\citenamefont {Gulans}, \citenamefont {Kontur}, \citenamefont {Meisenbichler}, \citenamefont {Nabok}, \citenamefont {Pavone}, \citenamefont {Rigamonti}, \citenamefont {Sagmeister}, \citenamefont {Werner},\ and\ \citenamefont {Draxl}}]{gulans_exciting_2014}%
  \BibitemOpen
  \bibfield  {author} {\bibinfo {author} {\bibfnamefont {A.}~\bibnamefont {Gulans}}, \bibinfo {author} {\bibfnamefont {S.}~\bibnamefont {Kontur}}, \bibinfo {author} {\bibfnamefont {C.}~\bibnamefont {Meisenbichler}}, \bibinfo {author} {\bibfnamefont {D.}~\bibnamefont {Nabok}}, \bibinfo {author} {\bibfnamefont {P.}~\bibnamefont {Pavone}}, \bibinfo {author} {\bibfnamefont {S.}~\bibnamefont {Rigamonti}}, \bibinfo {author} {\bibfnamefont {S.}~\bibnamefont {Sagmeister}}, \bibinfo {author} {\bibfnamefont {U.}~\bibnamefont {Werner}},\ and\ \bibinfo {author} {\bibfnamefont {C.}~\bibnamefont {Draxl}},\ }\bibfield  {title} {\bibinfo {title} {exciting: a full-potential all-electron package implementing density-functional theory and many-body perturbation theory},\ }\href {https://doi.org/10.1088/0953-8984/26/36/363202} {\bibfield  {journal} {\bibinfo  {journal} {J. Phys.: Condens. Matter}\ }\textbf {\bibinfo {volume} {26}},\ \bibinfo {pages} {363202} (\bibinfo {year} {2014})}\BibitemShut {NoStop}%
\bibitem [{\citenamefont {Draxl}\ and\ \citenamefont {Cocchi}(2017)}]{draxl_exciting_2017}%
  \BibitemOpen
  \bibfield  {author} {\bibinfo {author} {\bibfnamefont {C.}~\bibnamefont {Draxl}}\ and\ \bibinfo {author} {\bibfnamefont {C.}~\bibnamefont {Cocchi}},\ }\href {https://arxiv.org/abs/1709.02288} {\bibinfo {title} {Exciting core-level spectroscopy}} (\bibinfo {year} {2017}),\ \Eprint {https://arxiv.org/abs/1709.02288} {arXiv:1709.02288 [cond-mat.mtrl-sci]} \BibitemShut {NoStop}%
\bibitem [{\citenamefont {Vorwerk}\ \emph {et~al.}(2017)\citenamefont {Vorwerk}, \citenamefont {Cocchi},\ and\ \citenamefont {Draxl}}]{vorwerk_addressing_2017}%
  \BibitemOpen
  \bibfield  {author} {\bibinfo {author} {\bibfnamefont {C.}~\bibnamefont {Vorwerk}}, \bibinfo {author} {\bibfnamefont {C.}~\bibnamefont {Cocchi}},\ and\ \bibinfo {author} {\bibfnamefont {C.}~\bibnamefont {Draxl}},\ }\bibfield  {title} {\bibinfo {title} {Addressing electron-hole correlation in core excitations of solids: {An} all-electron many-body approach from first principles},\ }\href {https://doi.org/10.1103/PhysRevB.95.155121} {\bibfield  {journal} {\bibinfo  {journal} {Phys. Rev. B}\ }\textbf {\bibinfo {volume} {95}},\ \bibinfo {pages} {155121} (\bibinfo {year} {2017})}\BibitemShut {NoStop}%
\bibitem [{\citenamefont {Wang}\ \emph {et~al.}(2019)\citenamefont {Wang}, \citenamefont {Elliott}, \citenamefont {Sharma},\ and\ \citenamefont {Dewhurst}}]{wang_real_2019}%
  \BibitemOpen
  \bibfield  {author} {\bibinfo {author} {\bibfnamefont {C.-Y.}\ \bibnamefont {Wang}}, \bibinfo {author} {\bibfnamefont {P.}~\bibnamefont {Elliott}}, \bibinfo {author} {\bibfnamefont {S.}~\bibnamefont {Sharma}},\ and\ \bibinfo {author} {\bibfnamefont {J.~K.}\ \bibnamefont {Dewhurst}},\ }\bibfield  {title} {\bibinfo {title} {Real time scissor correction in {TD}-{DFT}},\ }\href {https://doi.org/10.1088/1361-648X/ab048a} {\bibfield  {journal} {\bibinfo  {journal} {J. Phys.: Condens. Matter}\ }\textbf {\bibinfo {volume} {31}},\ \bibinfo {pages} {214002} (\bibinfo {year} {2019})}\BibitemShut {NoStop}%
\bibitem [{\citenamefont {Zürch}\ \emph {et~al.}(2017)\citenamefont {Zürch}, \citenamefont {Chang}, \citenamefont {Borja}, \citenamefont {Kraus}, \citenamefont {Cushing}, \citenamefont {Gandman}, \citenamefont {Kaplan}, \citenamefont {Oh}, \citenamefont {Prell}, \citenamefont {Prendergast}, \citenamefont {Pemmaraju}, \citenamefont {Neumark},\ and\ \citenamefont {Leone}}]{zurch_direct_2017}%
  \BibitemOpen
  \bibfield  {author} {\bibinfo {author} {\bibfnamefont {M.}~\bibnamefont {Zürch}}, \bibinfo {author} {\bibfnamefont {H.-T.}\ \bibnamefont {Chang}}, \bibinfo {author} {\bibfnamefont {L.~J.}\ \bibnamefont {Borja}}, \bibinfo {author} {\bibfnamefont {P.~M.}\ \bibnamefont {Kraus}}, \bibinfo {author} {\bibfnamefont {S.~K.}\ \bibnamefont {Cushing}}, \bibinfo {author} {\bibfnamefont {A.}~\bibnamefont {Gandman}}, \bibinfo {author} {\bibfnamefont {C.~J.}\ \bibnamefont {Kaplan}}, \bibinfo {author} {\bibfnamefont {M.~H.}\ \bibnamefont {Oh}}, \bibinfo {author} {\bibfnamefont {J.~S.}\ \bibnamefont {Prell}}, \bibinfo {author} {\bibfnamefont {D.}~\bibnamefont {Prendergast}}, \bibinfo {author} {\bibfnamefont {C.~D.}\ \bibnamefont {Pemmaraju}}, \bibinfo {author} {\bibfnamefont {D.~M.}\ \bibnamefont {Neumark}},\ and\ \bibinfo {author} {\bibfnamefont {S.~R.}\ \bibnamefont {Leone}},\ }\bibfield  {title} {\bibinfo {title} {Direct and {Simultaneous} {Observation} of {Ultrafast} {Electron} and {Hole} {Dynamics} in {Germanium}},\
  }\href {https://doi.org/10.1038/ncomms15734} {\bibfield  {journal} {\bibinfo  {journal} {Nat. Commun.}\ }\textbf {\bibinfo {volume} {8}},\ \bibinfo {pages} {15734} (\bibinfo {year} {2017})}\BibitemShut {NoStop}%
\bibitem [{\citenamefont {Matthew}(1970)}]{matthew_temperature_1970}%
  \BibitemOpen
  \bibfield  {author} {\bibinfo {author} {\bibfnamefont {J.~A.~D.}\ \bibnamefont {Matthew}},\ }\bibfield  {title} {\bibinfo {title} {A temperature dependent contribution to {Auger} electron energy distributions},\ }\href {https://doi.org/10.1016/0039-6028(70)90219-0} {\bibfield  {journal} {\bibinfo  {journal} {Surf. Sci.}\ }\textbf {\bibinfo {volume} {20}},\ \bibinfo {pages} {183} (\bibinfo {year} {1970})}\BibitemShut {NoStop}%
\bibitem [{\citenamefont {Almbladh}(1977)}]{almbladh_effects_1977}%
  \BibitemOpen
  \bibfield  {author} {\bibinfo {author} {\bibfnamefont {C.-O.}\ \bibnamefont {Almbladh}},\ }\bibfield  {title} {\bibinfo {title} {Effects of incomplete phonon relaxation on x-ray emission edges in simple metals},\ }\href {https://doi.org/10.1103/PhysRevB.16.4343} {\bibfield  {journal} {\bibinfo  {journal} {Phys. Rev. B}\ }\textbf {\bibinfo {volume} {16}},\ \bibinfo {pages} {4343} (\bibinfo {year} {1977})}\BibitemShut {NoStop}%
\bibitem [{\citenamefont {Mahan}(1980)}]{mahan_photoemission_1980}%
  \BibitemOpen
  \bibfield  {author} {\bibinfo {author} {\bibfnamefont {G.~D.}\ \bibnamefont {Mahan}},\ }\bibfield  {title} {\bibinfo {title} {{Photoemission from alkali halides: Energies and line shapes}},\ }\href {https://doi.org/10.1103/PhysRevB.21.4791} {\bibfield  {journal} {\bibinfo  {journal} {Phys. Rev. B}\ }\textbf {\bibinfo {volume} {21}},\ \bibinfo {pages} {4791} (\bibinfo {year} {1980})}\BibitemShut {NoStop}%
\bibitem [{\citenamefont {Gallon}\ and\ \citenamefont {Matthew}(1970)}]{gallon_low_1970}%
  \BibitemOpen
  \bibfield  {author} {\bibinfo {author} {\bibfnamefont {T.~E.}\ \bibnamefont {Gallon}}\ and\ \bibinfo {author} {\bibfnamefont {J.~A.~D.}\ \bibnamefont {Matthew}},\ }\bibfield  {title} {\bibinfo {title} {{Low Energy Auger Emission from Lithium Fluoride}},\ }\href {https://doi.org/10.1002/pssb.19700410138} {\bibfield  {journal} {\bibinfo  {journal} {Phys. Stat. Sol. (b)}\ }\textbf {\bibinfo {volume} {41}},\ \bibinfo {pages} {343} (\bibinfo {year} {1970})}\BibitemShut {NoStop}%
\bibitem [{\citenamefont {Matthew}\ and\ \citenamefont {Komninos}(1975)}]{matthew_transition_1975}%
  \BibitemOpen
  \bibfield  {author} {\bibinfo {author} {\bibfnamefont {J.~A.~D.}\ \bibnamefont {Matthew}}\ and\ \bibinfo {author} {\bibfnamefont {Y.}~\bibnamefont {Komninos}},\ }\bibfield  {title} {\bibinfo {title} {{Transition rates for interatomic Auger processes}},\ }\href {https://doi.org/10.1016/0039-6028(75)90166-1} {\bibfield  {journal} {\bibinfo  {journal} {Surf. Sci.}\ }\textbf {\bibinfo {volume} {53}},\ \bibinfo {pages} {716} (\bibinfo {year} {1975})}\BibitemShut {NoStop}%
\bibitem [{\citenamefont {Hotokka}\ \emph {et~al.}(1984)\citenamefont {Hotokka}, \citenamefont {\AA{}gren}, \citenamefont {Aksela},\ and\ \citenamefont {Aksela}}]{hotokka_auger_1984}%
  \BibitemOpen
  \bibfield  {author} {\bibinfo {author} {\bibfnamefont {M.}~\bibnamefont {Hotokka}}, \bibinfo {author} {\bibfnamefont {H.}~\bibnamefont {\AA{}gren}}, \bibinfo {author} {\bibfnamefont {H.}~\bibnamefont {Aksela}},\ and\ \bibinfo {author} {\bibfnamefont {S.}~\bibnamefont {Aksela}},\ }\bibfield  {title} {\bibinfo {title} {{Auger spectrum of the LiF molecule}},\ }\href {https://doi.org/10.1103/PhysRevA.30.1855} {\bibfield  {journal} {\bibinfo  {journal} {Phys. Rev. A}\ }\textbf {\bibinfo {volume} {30}},\ \bibinfo {pages} {1855} (\bibinfo {year} {1984})}\BibitemShut {NoStop}%
\bibitem [{\citenamefont {Dolling}\ \emph {et~al.}(1968)\citenamefont {Dolling}, \citenamefont {Smith}, \citenamefont {Nicklow}, \citenamefont {Vijayaraghavan},\ and\ \citenamefont {Wilkinson}}]{dolling_lattice_1968}%
  \BibitemOpen
  \bibfield  {author} {\bibinfo {author} {\bibfnamefont {G.}~\bibnamefont {Dolling}}, \bibinfo {author} {\bibfnamefont {H.~G.}\ \bibnamefont {Smith}}, \bibinfo {author} {\bibfnamefont {R.~M.}\ \bibnamefont {Nicklow}}, \bibinfo {author} {\bibfnamefont {P.~R.}\ \bibnamefont {Vijayaraghavan}},\ and\ \bibinfo {author} {\bibfnamefont {M.~K.}\ \bibnamefont {Wilkinson}},\ }\bibfield  {title} {\bibinfo {title} {{Lattice Dynamics of Lithium Fluoride}},\ }\href {https://doi.org/10.1103/PhysRev.168.970} {\bibfield  {journal} {\bibinfo  {journal} {Phys. Rev.}\ }\textbf {\bibinfo {volume} {168}},\ \bibinfo {pages} {970} (\bibinfo {year} {1968})}\BibitemShut {NoStop}%
\bibitem [{\citenamefont {Willett-Gies}\ \emph {et~al.}(2015)\citenamefont {Willett-Gies}, \citenamefont {Nelson}, \citenamefont {Abdallah},\ and\ \citenamefont {Zollner}}]{willett-gies_two-phonon_2015}%
  \BibitemOpen
  \bibfield  {author} {\bibinfo {author} {\bibfnamefont {T.~I.}\ \bibnamefont {Willett-Gies}}, \bibinfo {author} {\bibfnamefont {C.~M.}\ \bibnamefont {Nelson}}, \bibinfo {author} {\bibfnamefont {L.~S.}\ \bibnamefont {Abdallah}},\ and\ \bibinfo {author} {\bibfnamefont {S.}~\bibnamefont {Zollner}},\ }\bibfield  {title} {\bibinfo {title} {{Two-phonon absorption in LiF and NiO from infrared ellipsometry}},\ }\href {https://doi.org/10.1116/1.4927159} {\bibfield  {journal} {\bibinfo  {journal} {J. Vac. Sci. Technol. A}\ }\textbf {\bibinfo {volume} {33}},\ \bibinfo {pages} {061202} (\bibinfo {year} {2015})}\BibitemShut {NoStop}%
\bibitem [{\citenamefont {Singh}\ and\ \citenamefont {Upadhyaya}(1972)}]{singh_crystal_1972}%
  \BibitemOpen
  \bibfield  {author} {\bibinfo {author} {\bibfnamefont {R.~K.}\ \bibnamefont {Singh}}\ and\ \bibinfo {author} {\bibfnamefont {K.~S.}\ \bibnamefont {Upadhyaya}},\ }\bibfield  {title} {\bibinfo {title} {{Crystal Dynamics of Magnesium Oxide}},\ }\href {https://doi.org/10.1103/PhysRevB.6.1589} {\bibfield  {journal} {\bibinfo  {journal} {Phys. Rev. B}\ }\textbf {\bibinfo {volume} {6}},\ \bibinfo {pages} {1589} (\bibinfo {year} {1972})}\BibitemShut {NoStop}%
\bibitem [{\citenamefont {Timmers}\ \emph {et~al.}(2017)\citenamefont {Timmers}, \citenamefont {Kobayashi}, \citenamefont {Chang}, \citenamefont {Reduzzi}, \citenamefont {Neumark},\ and\ \citenamefont {Leone}}]{timmers_generating_2017}%
  \BibitemOpen
  \bibfield  {author} {\bibinfo {author} {\bibfnamefont {H.}~\bibnamefont {Timmers}}, \bibinfo {author} {\bibfnamefont {Y.}~\bibnamefont {Kobayashi}}, \bibinfo {author} {\bibfnamefont {K.~F.}\ \bibnamefont {Chang}}, \bibinfo {author} {\bibfnamefont {M.}~\bibnamefont {Reduzzi}}, \bibinfo {author} {\bibfnamefont {D.~M.}\ \bibnamefont {Neumark}},\ and\ \bibinfo {author} {\bibfnamefont {S.~R.}\ \bibnamefont {Leone}},\ }\bibfield  {title} {\bibinfo {title} {{Generating high-contrast, near single-cycle waveforms with third-order dispersion compensation}},\ }\href {https://doi.org/10.1364/OL.42.000811} {\bibfield  {journal} {\bibinfo  {journal} {Opt. Lett.}\ }\textbf {\bibinfo {volume} {42}},\ \bibinfo {pages} {811} (\bibinfo {year} {2017})}\BibitemShut {NoStop}%
\bibitem [{\citenamefont {Gilbertson}\ \emph {et~al.}(2010)\citenamefont {Gilbertson}, \citenamefont {Chini}, \citenamefont {Feng}, \citenamefont {Khan}, \citenamefont {Wu},\ and\ \citenamefont {Chang}}]{gilbertson_monitoring_2010}%
  \BibitemOpen
  \bibfield  {author} {\bibinfo {author} {\bibfnamefont {S.}~\bibnamefont {Gilbertson}}, \bibinfo {author} {\bibfnamefont {M.}~\bibnamefont {Chini}}, \bibinfo {author} {\bibfnamefont {X.}~\bibnamefont {Feng}}, \bibinfo {author} {\bibfnamefont {S.}~\bibnamefont {Khan}}, \bibinfo {author} {\bibfnamefont {Y.}~\bibnamefont {Wu}},\ and\ \bibinfo {author} {\bibfnamefont {Z.}~\bibnamefont {Chang}},\ }\bibfield  {title} {\bibinfo {title} {{Monitoring and Controlling the Electron Dynamics in Helium with Isolated Attosecond Pulses}},\ }\href {https://doi.org/10.1103/PhysRevLett.105.263003} {\bibfield  {journal} {\bibinfo  {journal} {Phys. Rev. Lett.}\ }\textbf {\bibinfo {volume} {105}},\ \bibinfo {pages} {263003} (\bibinfo {year} {2010})}\BibitemShut {NoStop}%
\bibitem [{\citenamefont {Kaldun}\ \emph {et~al.}(2016)\citenamefont {Kaldun}, \citenamefont {Bl{\"a}ttermann}, \citenamefont {Stoo{\ss}}, \citenamefont {Donsa}, \citenamefont {Wei}, \citenamefont {Pazourek}, \citenamefont {Nagele}, \citenamefont {Ott}, \citenamefont {Lin}, \citenamefont {Burgd{\"o}rfer},\ and\ \citenamefont {Pfeifer}}]{kaldun_observing_2016}%
  \BibitemOpen
  \bibfield  {author} {\bibinfo {author} {\bibfnamefont {A.}~\bibnamefont {Kaldun}}, \bibinfo {author} {\bibfnamefont {A.}~\bibnamefont {Bl{\"a}ttermann}}, \bibinfo {author} {\bibfnamefont {V.}~\bibnamefont {Stoo{\ss}}}, \bibinfo {author} {\bibfnamefont {S.}~\bibnamefont {Donsa}}, \bibinfo {author} {\bibfnamefont {H.}~\bibnamefont {Wei}}, \bibinfo {author} {\bibfnamefont {R.}~\bibnamefont {Pazourek}}, \bibinfo {author} {\bibfnamefont {S.}~\bibnamefont {Nagele}}, \bibinfo {author} {\bibfnamefont {C.}~\bibnamefont {Ott}}, \bibinfo {author} {\bibfnamefont {C.~D.}\ \bibnamefont {Lin}}, \bibinfo {author} {\bibfnamefont {J.}~\bibnamefont {Burgd{\"o}rfer}},\ and\ \bibinfo {author} {\bibfnamefont {T.}~\bibnamefont {Pfeifer}},\ }\bibfield  {title} {\bibinfo {title} {{Observing the ultrafast buildup of a Fano resonance in the time domain}},\ }\href {https://doi.org/10.1126/science.aah6972} {\bibfield  {journal} {\bibinfo  {journal} {Science}\ }\textbf {\bibinfo {volume} {354}},\ \bibinfo {pages} {738} (\bibinfo {year}
  {2016})}\BibitemShut {NoStop}%
\bibitem [{\citenamefont {Mori-S\'anchez}\ \emph {et~al.}(2008)\citenamefont {Mori-S\'anchez}, \citenamefont {Cohen},\ and\ \citenamefont {Yang}}]{MoriSanchez2008}%
  \BibitemOpen
  \bibfield  {author} {\bibinfo {author} {\bibfnamefont {P.}~\bibnamefont {Mori-S\'anchez}}, \bibinfo {author} {\bibfnamefont {A.~J.}\ \bibnamefont {Cohen}},\ and\ \bibinfo {author} {\bibfnamefont {W.}~\bibnamefont {Yang}},\ }\bibfield  {title} {\bibinfo {title} {Localization and delocalization errors in density functional theory and implications for band-gap prediction},\ }\href {https://doi.org/10.1103/PhysRevLett.100.146401} {\bibfield  {journal} {\bibinfo  {journal} {Phys. Rev. Lett.}\ }\textbf {\bibinfo {volume} {100}},\ \bibinfo {pages} {146401} (\bibinfo {year} {2008})}\BibitemShut {NoStop}%
\bibitem [{\citenamefont {Perdew}\ and\ \citenamefont {Levy}(1983)}]{Perdew1983}%
  \BibitemOpen
  \bibfield  {author} {\bibinfo {author} {\bibfnamefont {J.~P.}\ \bibnamefont {Perdew}}\ and\ \bibinfo {author} {\bibfnamefont {M.}~\bibnamefont {Levy}},\ }\bibfield  {title} {\bibinfo {title} {Physical content of the exact kohn-sham orbital energies: Band gaps and derivative discontinuities},\ }\href {https://doi.org/10.1103/PhysRevLett.51.1884} {\bibfield  {journal} {\bibinfo  {journal} {Phys. Rev. Lett.}\ }\textbf {\bibinfo {volume} {51}},\ \bibinfo {pages} {1884} (\bibinfo {year} {1983})}\BibitemShut {NoStop}%
\bibitem [{\citenamefont {Sham}\ and\ \citenamefont {Schl\"uter}(1983)}]{Sham1983}%
  \BibitemOpen
  \bibfield  {author} {\bibinfo {author} {\bibfnamefont {L.~J.}\ \bibnamefont {Sham}}\ and\ \bibinfo {author} {\bibfnamefont {M.}~\bibnamefont {Schl\"uter}},\ }\bibfield  {title} {\bibinfo {title} {Density-functional theory of the energy gap},\ }\href {https://doi.org/10.1103/PhysRevLett.51.1888} {\bibfield  {journal} {\bibinfo  {journal} {Phys. Rev. Lett.}\ }\textbf {\bibinfo {volume} {51}},\ \bibinfo {pages} {1888} (\bibinfo {year} {1983})}\BibitemShut {NoStop}%
\bibitem [{\citenamefont {Géneaux}\ \emph {et~al.}(2021)\citenamefont {Géneaux}, \citenamefont {Chang}, \citenamefont {Schwartzberg},\ and\ \citenamefont {Marroux}}]{geneaux_source_2021}%
  \BibitemOpen
  \bibfield  {author} {\bibinfo {author} {\bibfnamefont {R.}~\bibnamefont {Géneaux}}, \bibinfo {author} {\bibfnamefont {H.-T.}\ \bibnamefont {Chang}}, \bibinfo {author} {\bibfnamefont {A.~M.}\ \bibnamefont {Schwartzberg}},\ and\ \bibinfo {author} {\bibfnamefont {H.~J.~B.}\ \bibnamefont {Marroux}},\ }\bibfield  {title} {\bibinfo {title} {Source noise suppression in attosecond transient absorption spectroscopy by edge-pixel referencing},\ }\href {https://doi.org/10.1364/OE.412117} {\bibfield  {journal} {\bibinfo  {journal} {Opt. Express}\ }\textbf {\bibinfo {volume} {29}},\ \bibinfo {pages} {951} (\bibinfo {year} {2021})}\BibitemShut {NoStop}%
\end{thebibliography}%

\end{document}